\documentclass[12pt]{article}
\title{
THEORY OF LIGHT QUARKS IN THE CONFINING VACUUM
}
\author{Yu.A.Simonov\\
Institute of Theoretical and Experimental Physics\\ 117118,
Moscow, B.Cheremushkinskaya 25, Russia}
\newcommand{\be}{\begin{equation}}
\newcommand{\ee}{\end{equation}}
\def\la{\mathrel{\mathpalette\fun <}}
\def\ga{\mathrel{\mathpalette\fun >}}
\def\fun#1#2{\lower3.6pt\vbox{\baselineskip0pt\lineskip.9pt
\ialign{$\mathsurround=0pt#1\hfil
##\hfil$\crcr#2\crcr\sim\crcr}}}

\begin{document}
\maketitle

\large

\begin{abstract}

The light quark propagation in the confining vacuum, described by an
(infinite) set of gauge--field vacuum correlators, is studied in
detail. To keep gauge invariance at each step the system of light
quark and a heavy antiquark is considered, and the nonlinear
equations are written explicitly for the quark propagator in the
limit of large $N_c$. For the lowest (Gaussian) correlator the system
is studied in different approximations, and the
relativistic WKB method is used to demonstrate the scalar
confining interaction of light quarks, which implies
chiral symmetry breaking. The chiral condensate is
estimated by the WKB method, and connection to the
density of global zero modes is clarified. The higher
even order correlators are shown to yield the same
properties of scalar  confining   interaction for light
quarks. No attempt  was  made to solve the obtained
nonlinear equations numerically, but  the qualitative
conclusion of connection between confinement and chiral
symmetry breaking is drawn, and an estimate of the chiral condensate
is performed.

\end{abstract}

\section{Introduction}

The light quark propagation in the QCD vacuum displays two phenomena:
confinement and chiral symmetry breaking (CSB). The first one can be
most clearly studied in the example of heavy quarks, where the linear
confining potential (equivalent to  the area law of the Wilson loop)
is a good order parameter in the absence of  dynamical quarks.
Recently the new vacuum correlator method was introduced [1] which
successfully describes dynamics of confinement, and the connection
with the dual Meissner effect can be clearly seen (for a recent
review see [2]).

The confining quark dynamics for massive quarks was  formulated
in the vacuum correlator method,  and the method allows to take into
account spin degrees of freedom [3-5] but only as a perturbation  in
powers of $1/M$. E.g. one can extend the method to
effectively calculate spin-dependent contributions to the mass of
$\rho$ meson, but one fails in the case of pion.

This difficulty can be traced back to the lack of CSB in the
Feynman--Schwinger representation for the quark propagator with spin
[6].

To treat CSB one can use the most popular way -- to consider a gas or
liquid of topological charges in the vacuum. For the case of
instantons it was argued that CSB occurs ( at least for large
$N_c$) for any instanton density ([7], for a recent review see [8]),
a similar result holds for dyonic gas [9].

Therefore a logical way of creating a natural QCD environments for
light quarks in this approach consisted of placing topological
charges inside a confining vacuum, described by vacuum correlators.
It was shown in [10] that such a construction indeed provides CSB
together with confinement in a gauge--invariant way, and moreover
instantons are stabilized at large distances  due to confinement
[11].  In particular one could visualize in such vacuum the double
nature of pion as a $q\bar q$ bound state and as a Nambu--Goldstone
particle in one and the same physical Green's function.

There are visible defects in such construction however. Firstly, it
looks artificial to superimpose  instantons in the confining
background. Secondly, there is no explanation of why confinement and
CSB occur together in the confining phase and why they disappear
simultaneously above $T_c$, as it was shown repeatedly on the lattice
[12].

There is another type of approach to CSB in connection with
confinement [13], where Dyson--Schwinger equations are used for the
isolated quark propagator with selfinteraction via modified gluon
exchange However the system under study is not gauge invariant and
physically there one does not take into account the QCD string,
connecting the quark to antiquark.

 In this paper we choose therefore
another way.  We start with the QCD Lagrangian and derive from that
the effective Lagrangian of light quarks, assuming that certain gluon
field correlators are nonzero, which are known to yield confinement
(i.e.  linear potential) for static quarks. It is not clear from the
beginning what will happen for light quarks ( with vanishing mass)
and whether they would be confined at all.

Our main concern in what follows is to keep Lorentz and gauge
invariance. The effective quark interaction is nonlocal and to have
gauge-invariant equation one should consider $q\bar q$ Green's
functions.

The simplest setting for which confinement and CSB can  be studied in
the gauge--invariant way, is the problem of a light  quark
propagation in the field of the static antiquark. To simplify matter
we start with the Gaussian correlators for gluon fields and derive
the selfconsistent  equations for the light quark propagator (with a
string effectively connecting it to the static source). We show that
CSB occurs due to the string (linear confinement), which shows up in
the fact that effective interaction becomes Lorentz scalar. We also
check the limit of heavy quark mass and  demonstrate the usual
linear potential in this case. As another evidence of CSB the
chiral condensate is computed and shown to be nonzero.

Those results were derived actually for the case of one light
quark (quenching approximation or $N_c\to \infty$ ) and when only
bilocal field correlators  are kept nonzero (Gaussian vacuum
approximation). To treat the case of two and more flavours the
corresponding term in the effective  Lagrangian may be studied, and
one is naturally led to the equation for the Green's  function of two
light quarks. The qualitative discussion of this equations leads to
the same conclusion as in the case of one light quark--namely, the
Gaussian field correlators which ensure linear confinement for
static quarks, also yield  for two light quarks interaction
kernels growing linearly at large distances and produce CSB at the
same time.

Finally one could ask the question: what the effect of higher (non
Gaussian) field   correlators will be on the stated  above results.
It is argued below
 that  higher correlators
  of even order
   bring about the same results as for the
 Gaussian correlators. One may then ask about the effect of the
 infinite sum of correlators, as it is in reality. The answer is that
 when higher order correlators only renormalize the string tension
 and do not make it vanish, all the main features of CSB hold, in the
 opposite case, however, additional investigation is needed.

The paper is organized as follows. In the second section the general
form of the effective Lagrangian for light quarks is  given, and the
equation for the Green's function of one light quark in the field of
the static source is derived. In the third section this equation is
studied and the conditions for the CSB are derived, which are
satisfied in the presence of the Gaussian field correlators.

The fourth section is devoted to the study of equations for the quark
propagator in the limit of small correltion length $T_g$, when the
kernel of the equation becomes quasilocal.

In the fifth   section the powerful  relativistic WKB methods are
used to solve equations and calculate the kernel at large distances.
It enables one to calculate the chiral condensate in the limit of
small $T_g$. This is done in section 6. One  discovers there that
parametrically the chiral condensate  is proportional to
$\sigma/T_g$, where $\sigma$ is the string tension, and therefore
diverges in the "string limit", when $\sigma$ is constant and $T_g$
vanishes.

Therefore some additional care is needed to get the chiral
condensate, and exact equations are written explaining these
difficulties.

The section 7 connects the chiral condensate to quark zero modes and
to the field correlators. It is shown, that confinement occurs due to
the correlator, which is expressed via monopole currents and the
latter are connected to zero modes.

The contribution of higher field correlators to the kernel of the
equations, and finally to confinement and CSB is discussed in section
8. Discussion and prospectives are presented in section 9.

The paper contains 5  appendices. Appendix 1 is devoted to the gauge
-invariant derivation of the effective Lagrangian, used in the main
text. In Appendix 2 another term in the Gaussian correlator is
studied which does not ensure confinement, and it is shown that it
also does not ensure CSB. Properties of the kernel, containing the
confining correlator $D(x)$, are investigated in detail in Appendix
3. Expansion and corrections to the kernel in powers of $T_g$ are
given in Appendix 4. Finally in Appendix 5 the limit of large mass
$m$ is investigated in detail.

\section{Derivation of the effective Lagrangian for the light quark}

To make discussion of confinement and chiral symmetry breaking fully
gauge invariant, we consider the gauge invariant physical amplitude
-- the Green's function $S$ of a light quark of mass $m$ in the
field of a static antiquark placed at the origin. The propagator for
the latter can be taken as (we consider below Euclidean space-time).
\be
S_{\bar Q}(T) \equiv \frac{1+\gamma_4}{2}P~exp~ig\int^T_0A_4(\vec
r=0,\tau) d\tau
\ee

The Green's function $S_{q\bar Q}$ can be written as an integral
\be
S_{q\bar Q}(x,y)=\frac{1}{N}\int
D{\psi}D{\psi^+}DA
e^{-\int\frac{F_{\mu\nu}^2}{4} d^4x
-\int{\psi^+}(-i\hat \partial -im -\hat A)\psi d^4x}
\psi^+(x) S_{\bar Q} (x,y)\psi(y)
\ee

Since $S$ is gauge invariant, one can choose any convenient gauge for
$A_{\mu}$ and our  choice will be the modified Fock--Schwinger gauge,
introduced in [14], namely:
\be
A_4(\vec r=0,\tau)=0,~~r_iA_i(\vec r, \tau) =0,~~i=1,2,3.
\ee

In this gauge $G_{\bar Q}$ reduces to the factor
$\frac{1+\gamma_4}{2}$ and one can now consider the
integration over $DA$ in (2) as a statistical averaging
process and use the cluster expansion [15],
 $a,b$~-~ color~indices
 $$
  \int DA
e^{\int\psi^+(x)\hat A(x)\psi(x)d^4x-\int\frac{1}{4}
F^2_{\mu\nu}d^4x}=<exp \psi^+_a(x)\hat A_{ab}(x)\psi_b(x) d^4x>
 $$
\be
=exp\{\int d^4x\psi^+(x)\gamma_{\mu}<A_{\mu}>\psi(x)+
\ee
$$+\frac{1}{2}\int
dxdy\psi^+(x)\gamma_{\mu}\psi(x)
\psi^+(y)\gamma_{\nu}\psi(y)<A_{\mu}(x)A_{\nu}(y)>+...\}
$$

We have denoted the higher order cumulant contribution in (4) as
$...$ and shall disregard it for the most part of the paper,
coming back to it in section 8. The first term $<A_{\mu}>$ vanishes
due to the gauge and Lorentz invariance of the vacuum, while the
second can be expressed through the field strength correlators
$<F_{\mu\nu}(u)F_{\lambda\sigma}(u')>$ in the gauge (3) as follows
\be
A_4(\vec z,z_4)=\int^{z_i}_0du_iF_{i4}(u_i,z_4)
\ee
\be
A_k(\vec z,z_4)=\int^{z_i}_0 \alpha (u)du_iF_{ik}(u,z_4),~~
\alpha(u)=\frac{u_i}{z_i}
\ee

One can  easily see that the representation (5-6) satisfies condition
(3). Using (5,6) one can rewrite the average $<AA>$ in (4) as
$$
<A^{ab}_{\mu}(z)A^{cd}_{\mu'}(w)>=\frac{\delta_{bc}\delta_{ad}}{N_c}
\int^z_0 du_i\alpha_{\mu}(u)\int^w_0 du'_{i'}\alpha_{\mu'}(u')\times
  $$
  \be
  \times \frac{tr}{N_c} <F_{i\mu}(u)F_{i'\mu'}(u')>
  \ee
  where $ab,cd$ are fundamental color indices and we have defined
  $\alpha_4(u)\equiv 1,\alpha_k(u)\equiv \alpha(u), k=1,2,3.$
  The gauge--nonivariance of the correlator\\ $<FF>$ in (7) is only
  apparent and one can introduce a factor, equal to unity in the
  gauge (3), which makes the correlator explicitly gauge--invariant,
  namely
  \be
  <F(u)F(u')>=<F(\vec u, z_4)\Psi F(\vec u', w_4) \Psi^+>
  \ee
  where $\Psi$ is the product of 3 parallel transporters
  \be
  \Psi\equiv \Phi(\vec u z_4; oz_4)\Phi(oz_4,ow_4) \Phi(ow_4; \vec
  u'w_4)
  \ee
  and
$$
  \Phi(P,P')=P~exp~ig \int^P_{P'} dv_{\mu}A_{\mu}(v)
$$
   where the path-ordered contour integral $\Phi$ is taken along the
   straight line connecting the points $P$ and $P'$.

   Since in the gauge (3) one has $\Psi=\Psi^+\equiv 1$, we shall
   below omit those factors. For the correlator $<FF>$ one can use
   the parametrization suggested in the second entry of  [1]
   \be
   g^2<F_{i\mu}(u)F_{i'\mu'}(u')>_{ab}=\delta_{ab}(\delta_{ii'}\delta_{\mu\mu'}-
   \delta_{i\mu'}\delta_{i'\mu}) D(u-u')+\Delta^{(1)}
   \ee
   where $\Delta^{(1)}$ is proportional to a full derivative, its
   exact form is given in Appendix 2.

   In what follows we shall consider mostly the term $D$ in (10)
   since it contributes to the string tension, while $\Delta^{(1)}$
   does not. Namely using (10) it was obtained in [1] that the string
   tension $\sigma$ -- the coefficient in the area law of the Wilson
   loop, $<W(C)>=exp (-\sigma area)$ is equal to
   \be
   \sigma = \frac{1}{2} \int^{\infty}_{ -\infty} d^2uD(u)
   \ee

   Thus $\sigma$ characterizes the confinement of static quarks,
   and our goal is to understand how the dynamics of light quarks is
   expressible through $\sigma$ and whether $\sigma$ correlates with
   CSB.

The role of $\Delta^{(1)}$ is clarified later in section 3 when we
discuss  the deconfinement phase transition, with most details
contained in the Appendix 2.

Keeping only the term $D$ in (10) and neglecting higher order
correlators like $<AAA>$, one obtains in (4) the following effective
Lagrangian for the light quark:
$$
{\cal{L}}_{eff}(\psi^+\psi)=\int \psi^+(x)(-i\hat \partial -im )
\psi(x)d^4x +
$$
\be
\frac{1}{2N_c}\int
d^4xd^4y(\psi^+_a(x)\gamma_{\mu}\psi_b(x))(\psi^+_b(y)
\gamma_{\mu'}\psi_a(y))\times
\ee
$$
\times
(\delta_{\mu\mu'}\delta_{ii'}-
\delta_{i\mu'}\delta_{i'\mu})J^{\mu\mu'}_{ii'}
(x,y)
$$
where we have defined
\be
J^{\mu\mu'}_{ii'}(z,w)=\int^z_0 du_i\alpha_{\mu}(u)\int^w_0du'_{i'}
\alpha_{\mu'}(u')D(u-u')
\ee
and $\alpha_{\mu}=1$ for $\mu =4$,
and $\alpha_{\mu}=
\alpha (u)$
 for $\mu =1,2,3$.

The superscripts $(\mu,\mu')$ in $J$ in (13) enter at zero
temperature only through $\alpha_{\mu},\alpha_{\mu'}$ and will be
important for us in section 3, when we discuss the deconfinement
phase transition.

In what follows we disregard the perturbative contributions to
${\cal{L}}_{eff}$, since they have nothing to do with CSB. The mass
$m$  is supposed to be defined at the typical hadronic scale of 1
GeV, and we shall not be interested in its evolution to lower
scales.

From the effective Lagrangian (12) one can easily derive the equation
of Dyson-Schwinger type for the selfenergy part, which we shall
denote by $M$ and  the $q\bar Q$ Green's function $S$. This  is done
in the same way, as in the NJL model [16], since the structure of the
Lagrangian (12) is similar to that of  NJL however nonlocal.

The main essential difference is the presence of the string ,
connecting the light quark to the static source, this part is
concealed in $J$, Eq.(13) and therefore the selfenergy part $M$ is
actually not the set of the one--particle--irreducible diagrams, but
rather the $\bar q Q$ interaction kernel.
In what follows we shall replace in (2) the factor $G_{\bar Q}$
by unity and the resulting Green's function will be denoted $S$.

In the configuration space the equations for $M$ and $S$ are readily
obtained from (12) noting that in the mean--field approximation one
has to replace a pair of $\psi,\psi^+$ operators in (12) as
\be
\psi_b(x)\psi^+_b(y)\to<\psi_b(x)\psi^+_b(y)>=N_cS(x,y),
\ee
and finally one obtains
\be
iM(z,w)=J_{ik}^{\mu\mu}(z,w)\gamma_{\mu}S(z,w)\gamma_{\mu}\delta_{ik}
-J^{ik}_{ik}\gamma_kS(z,w)\gamma_i,
\ee
\be
(-i\hat{\partial}_z-im)S(z,w)-i\int M(z,z')S(z', w)d^4z'=
\delta^{(4)}(z-w)
\ee

The system of equations (15-16) defines unambiguously both the
interaction kernel $M$ and the Green's function $S$. One should
stress at this point again that both $S$ and $M$ are not the
one-particle operators but rather two--particle operators, with the
role of the second particle played by the static source. It is due to
this property, that $S$ and $M$ are gauge invariant operators, which
is very important to take confinement into account properly. Had we
worked with one--particle operators, as is the habit in QED and
sometimes also in QCD,  then we would immediately loose the gauge
invariance and the string, and hence confinement.

\section{Properties of the selfconsistent solutions}

To study equations (15-16) it is convenient to go over to the
momentum representation. Since however the kernel $J(z,w)$  is not
decreasing at large distances at $|\vec z|\sim|\vec w|\to \infty$,
we shall use at the intermediate stage the cut-off factor,
multiplying $J(z,w)$ by the factor $exp(-\alpha^2(\vec w^2+\vec
z^2))$, and letting finally $\alpha$ to go to zero. Being initially a
purely technical device, this trick appears to be of more significant
contents, since it will automatically separate dynamics of large
distances $(r\sim \alpha^{-1/2})$ and relatively small distances
(independent of $\alpha$).

To make our expressions more transparent, we also assume for the
correlator $D(u)$ in (13) the Gaussian form
\be
D(u)=D(0) e^{-\frac{u^2}{4T^2_g}},
\ee
We note now that due to (5),(6) the integration in (13) is done
at the constant value of
the Euclidean time component. Therefore it is convenient to
treat the 4-th component separately,  writing
\be
M(z,w)=\int M(p_4,\vec p,\vec p') e^{ip_4(z_4-w_4)+i\vec p\vec
z+i\vec p'\vec w}
\frac{dp_4d\vec pd\vec p'}{(2\pi)^7}
\ee
\be
S(z,w)=\int S(p_4,\vec p,\vec p') e^{ip_4(z_4-w_4)+i\vec p\vec
z+i\vec p'\vec w}
\frac{dp_4d\vec pd\vec p'}{(2\pi)^7}
\ee
and equations (15-16) have the form
$$
iM(p_4,\vec p,\vec p')=\int \frac{dp'_4d\vec qd\vec
q'}{(2\pi)^7}e^{-(p_4-p'_4)^2T_g^2}\times
$$
\be
\times
[J_{ii}^{\mu\mu}(\vec q, \vec q')\gamma_{\mu} S (p'_4, \vec p-\vec q,
\vec p'-\vec q')\gamma_{\mu}-
\ee
$$
-J^{ik}_{ik}(\vec q, \vec q')\gamma_k S (p'_4,\vec p-\vec q,\vec
p'-\vec q')\gamma_i]
$$
$$
(\hat p_4+\hat p-im) S(p_4,\vec p,\vec p)-i\int \frac{d^3\vec
q}{(2\pi)^3} M(p_4,\vec p,\vec q) S(p_4,-\vec q, \vec p')=
$$
\be
=(2\pi)^3\delta (\vec p+\vec p')
\ee
where we have defined
\be
J_{ik}^{\mu\mu'}(\vec q,\vec q')=\int\frac{\sigma}{\sqrt{\pi} T_g}
e^{-i\vec q \vec z- i\vec q'\vec w-\alpha (\vec z^2+\vec w^2)}
z_iw_kK_{\mu\mu'}(\vec z,\vec w)d\vec z d\vec w
\ee
and
\be
K_{\mu\mu'}(\vec z,\vec w)=\int^1_0 dt\int^1_0
dt'\alpha_{\mu}(t)\alpha_{\mu'}(t')e^{-\frac{(\vec z t-\vec w
t')^2}{4T_g^2}}
\ee
with
$\alpha_4(t)=1, \alpha_i(t)=t, i=1,2,3.$

In derivation of (22) we have used (11) to express $D(0)$ through
$\sigma$.

The integral over $d\vec z d\vec w$ in (22) can be done easily.

Next we introduce dimensionless momenta $Q,P$ etc. instead of $q,p$
as follows
\be
Q_i=\frac{q_i}{\sqrt{\alpha}},
P_i=\frac{p_i}{\sqrt{\alpha}},
P_4=\frac{p_4}{\sqrt{\alpha}},
\ee
Similarly one introduces in (23) dimensionless variables $\tau,\tau'$
instead of $tt'$ connected through
\be
(t,t')=T_g\sqrt{\alpha}(\tau,\tau')
\ee
Now in the kernel $J_{ik}$ the $\alpha$ dependence can be explicitly
written as
\be
J_{ik}^{44}=\frac{\sigma T_g}{\alpha^3} f^{44}_{ik}(Q,Q')
\ee
\be
J_{ik}^{ik}=\frac{\sigma T_g^3}{\alpha^2} f^{ik}_{ik}(Q,Q')
\ee
 here $f_{ik}$ are dimensionless functions of order one, when their
 dimensionless arguments are also of the order of one, fast
 decreasing at infinity of $Q,Q'$ and finite at small $Q,Q'$.
  The important fact about $f_{ik}$ is that they do not depend on
 $\alpha$ in the limit $T_g\sqrt{\alpha}\to 0$, which is solely of
 physical interest.

 For  small $Q,Q'$ one has
 \be
 f_{ik}^{44}(0,0) = \frac{2}{3}\delta_{ik} \pi^{5/2}
 \ee
 From  the dimensional analysis of equations (20),(21) one can derive
 the "$\alpha$ -- dimensionality" of $M$ and $S$, namely:
 \be
 M(p_4,\bar p,\bar p')= \alpha^{-3/2}\tilde M(P_4,P,P')
 \ee
 \be
 S(p_4,\vec p,\vec p')= \alpha^{-2}\tilde S(P_4,P,P')
 \ee
 Insertion of (29),(30), (26), (27) into (20),(21) yields
 dimensionless equations for $\tilde M$ and $\tilde S$
 \be
 (P_4\gamma_4+P_i\gamma_i-\frac{im}{\sqrt{\alpha}})
 \tilde S(P_4,P,P')-
 \ee
 $$
 - i\int \frac{d^3Q}{(2\pi)^3}\frac{\tilde M(P_4,P,Q)}{\sqrt{\alpha}}
 \tilde S(P_4,-Q,P')=
 (2\pi)^3\delta^{(3)}(P+P')
 $$
 $$
 i\tilde M(P_4,P,P')= \sigma T_g\int \frac{d P'_4}{2\pi}
 e^{-(P_4-P'_4)^2\alpha T_g^2}
 \frac{d^3Qd^3 Q'}{(2\pi)^6}
 \{ f^{44}_{ik} (Q,Q')\times
 $$
 $$
 \times \gamma_4\tilde S (P'_4,P-Q,P'-Q')\gamma_4 + \alpha T_g^2
 f^{ii}_{ii} (Q,Q') \gamma_i\tilde S (P'_4, P-Q, P'-Q') \gamma_i
 $$
 \be
 -\alpha T_g^2f_{ik}^{ik}(Q,Q') \gamma_k\tilde S(P'_4,P-Q, P'-Q')
 \gamma_i\}
 \ee
 In the rest of this section we shall analyze the system of equations
 (31), (32) in the physical limit, when $\alpha T_g^2\to 0$, while
 $\sigma T_g$ is kept fixed.

 To understand better properties of solutions of (31),(32) it is
 convenient to consider first the limit of large mass $m$. In this
 case one expects to get the usual Lippman -- Schwinger equation
 corresponding to the linear potential between two heavy quarks.

 The limit of the large mass $m$ has its own peculiarities which can
 be easier seen in the configuration space. Hence we come back to
 equation (15), where one should insert the large--mass propagator
 $S_m(z,w)$
 \be
 S_m(z_4-w_4, \vec z,\vec w)=\int e^{ip_4(z_4-w_4)+i\vec p\vec
 z+i\vec p' \vec w}
  \frac{dp_4}{2\pi}
  \frac{d\vec p d\vec
 p'}{(2\pi)^6} \frac{(2\pi)^3\delta(\vec p+\vec
 p')(p_4\gamma_4+im)}{p_4^2+m^2}
  \ee
   obtained in the
  lowest  order in $(\frac{M}{m})$ from (21).  Thus one has the usual
  form for $h_4\geq 0$
   \be
    S_m(h_4,\vec z,\vec
  w)=\frac{i(1+\gamma_4)}{2}\theta(h_4 ) \delta^{(3)} (\vec z-\vec
  w)
  \ee
   and we have defined $h_4\equiv z_4-w_4$ and have factored
  out the term $exp (-mh_4)$, which contributes to the total factor
  $exp(-mT)$ of the overall $q\bar Q$ Green's function.

  Now we introduce $S_m$ or the r.h.s. of (15)  and compute the
  kernel $J_{ik}$ given in (13). One has
  \be
  J_{ik}^{\mu\mu}(h_4,\vec z,\vec z)= D(0) e^{-\frac{h_4^2}{ 4
  T_g^2}} \int^1_0\int^1_0 \alpha_{\mu}(t)\alpha_{\mu} (t') dt dt'
  e^{- \frac{\vec z^2}{4T_g^2}(t-t')^2} z_iz_k
  \ee
   At large $|\vec
  z|$ one obtains
  \be
  J_{ik}^{\mu\mu}(h_4,\vec z,\vec z) = D(0)
  e^{-h^2_4}{4T_g^2} \sqrt{\pi} \frac{2Tg}{|\vec z|} z_iz_k \left (
  \begin{array}{ll}
  1,&\mu=4\\
  \frac{1}{3},&\mu=1,2,3
  \end{array} \right )
  \ee

  Expressing now $D(0)$ through the string tension $\sigma$ via (11)
  one obtains for  $M(z,w)$;
  \be
  M(h_4,\vec z, \vec w)=  e^{-\frac{h^2_4}{4Tg^2}}D(0) \sqrt{\pi}
  T_g|\vec z|(2+\frac{\gamma_4-1}{3})\delta^{(3)}(\vec z
  -\vec w) \theta (h_4)
  \ee
  The equivalent static potential $U(\vec z,\vec w)$ obtains when one
  integrates over the relative time $h_4$ in the limit $T_g\to 0$, and
  keeps only the $(+,+ )$ component in Lorentz indices (or
  equivalently puts $\gamma_4=1$)
  \be
  U_{++}(\vec z,\vec w) =\int  dh_4 M(h_4,\vec z, \vec w) =\sigma
  |\vec z| \delta^{(3)} (\vec z-\vec w)
  \ee
  Thus one indeed obtains the standard linear  potential for the
  heavy quark, which is Lorentz scalar.

  We now turn back to the equations (31),(32) and look for the limit
  $\alpha \to 0$. To this end we remark in (32) that while $P,P'Q,Q'$
  are confined to the finite limits independent on $\alpha$, $P_4$
  and $P'_4$ are not and can be of the order  of $\alpha^{-1/2}$,
  therefore one can neglect $P_i\gamma_i$ on the l.h.s. of (31), thus
  writing it as
  \be
  \tilde S (P_4,P,P')=
  \frac{P_4\gamma_4+\frac{i(m+\tilde M)}{\sqrt{\alpha}}}
  {P_4^2+\frac{(m+\tilde M)^2}{\alpha}}
  (2\pi)^3\delta^{(3)}(\vec P+\vec P')
  \ee
  Inserting this in (32) and integrating over $dP_4$ in the limit
  $\alpha T^2_g \to 0$ one obtains
  \be
  i\tilde M(P_4,P,P')= \frac{i\sigma T_g}{2}
  \int \frac{d^3Qd^3Q'}{(2\pi)^6}(2\pi)^3\delta ^{(3)}(P+P'-Q-Q')
  \ee
  $$
  [(m+\tilde M)^2]^{-1/2} (m+\tilde M) f_{ii}^{44}(Q,Q')
  $$
  Here $\tilde M$ on the r.h.s. is actually an operator and its
  dependence on momenta in the integrand should be properly written,
  as well as the definition of the power $(-1/2)$.
  We would like to make several comments on Eq.(40).

  i) First of all one can see that the r.h.s. of (40) does not depend
  on $\alpha$, so that all expressions can be considered as in the
  physical limit. To see it more clearly, one can use (30) to express
  the physical Green's function $S$ through the mass $\tilde M$ as
  follows
  \be
  S(p_4,\bar p,\bar p')=\frac{\hat p_4+i\tilde M}{p_4^2+\tilde
  M^2}(2\pi)^3\delta^{(3)}(\vec p+\vec p')
  \ee
  and one can see that $\tilde M$ enters the physical propagator, and
  is the real physical quantity.

  ii) Secondly, only the $A_4$ component (which generates
  $f^{44}_{ii})$ contributes in the limit $\alpha T^2_g \to 0$, while
  $A_i,i=1,2,3$ give vanishing contribution.  In terms of field
  strengths it means that only color-electric correlators $<E_iE_j>$
  contribute to $\tilde M$ in the physical limit  $\alpha \to 0$,
  while color--magnetic do not. This fact seems to be  connected to
  our choice of gauge, since in this  gauge $\alpha_{\mu=4}=1$, while
  $\alpha_{\mu=i}=t$  and the latter yield suppression of the
  contribution of $A_i,$ at large distances. However, it is clear
  that the gauge corresponds to the physical situation, when the
  static  quark has the world-line along the 4-th coordinate, and
  therefore the string which is formed between the light quark and
  the static one,   is evolving alone the 4-th coordinate, hence it is
  done by the electric field.  In the last section we consider this
  point more closely, since it explains why CSB disappears together
  with  electric correlators $D(x)(10)
  $ at the deconfinement
  phase transition.

  iii)  The most important feature of (40) is that it exhibits a
  finite scalar solution for $\tilde M$ even in the limit of
  vanishing quark mass $m$. This solution exists for any finite value
  of $\sigma T_g$, and is the consequence of behaviour at large
  distances  (corresponding to the limit $\alpha \to 0$).

  For less singular behaviour  of the interaction
  kernel $J_{ik}$ at large distances, such as (27) instead of (26),
  the selfconsistent solution (40) does not exist.

  iv) In (40) we have retained only terms of the correlator $D$,
  neglecting $D_1$. The role of $D_1$ is discussed in Appendix 2,
  where we show that it does not give selfconsistent solutions
  in contrast  to the case of $D$.

\section{The local limit of selfonsistent solutions. Confinement and
CSB for  light quarks}

Our  bassic  equations (15),(16) are nonlocal in time because of the
integral over $dz'_4$ in (16). This nonlocality and the parameter
which it governs can be handled most  easily, when one uses instead
of $M(z,z')$, $S(z,z')$ the  Fourier  transforms.

\be
S(z_4-z'_4,\vec z,\vec z')= \int e^{ip_4(z_4-z'_4)}
S(p_4,\vec z, \vec z') \frac{dp_4}{2\pi}
\ee
and the same for $M(z,z')$. Then instead of (15),(16) one obtains  a
system of equations
\be
(\hat p_4-i\hat \partial_z-im)S(p_4, \vec z, \vec w) - i\int
M(p_4,\vec z, \vec z') S(p_4, \vec z', \vec w) d \vec z'=
\delta^{(3)}(\vec z - \vec w)
\ee
$$
iM(p_4, \vec z, \vec w)= 2\sqrt{\pi}T_g \int \frac{dp'_4}{2\pi}
e^{-(p_4-p'_4)^2 T_g^2}\times
 $$
 \be
 \times [J_{ik}^{\mu\mu}(\vec z, \vec w)\gamma_{\mu}S(p'_4,\vec z,
 \vec w)\gamma_{\mu}\delta_{ik}-J^{ik}_{ik}\gamma_kS\gamma_i]
 \ee
 where $\hat \partial_z=\gamma_i\frac{\partial}{\partial z_i}$, $J$
 is defined in (13) and we have factored out the time--dependent
 exponent of $D(u)$, using the representation (17). (For any form of
 correlator $D(u)$ the main result below, Eq. () remains true, but
 the corrections to it are dependent on the shape of $D(u)$ and are
 displayed in Appendix 4).

 All dependence of $M$ on $p_4$ as can be seen in (44) is due to the
 factor $exp [-(p_4-p'_4)^2T_g^2]$    and disappears in the limit when
 $T_g$ goes to zero, while the string tension $\sigma\sim D(0)T^2_g$
 is kept  fixed. This limit can be called the string limit of QCD,
 and we shall study its consequences for equations (43),(44) in this
 section, while in the appendix 4 corrections to this limit are
 considered.

 So in the string limit, with $M$ independent of $p_4$, let us
 consider the hermitian Hamiltonian
 \be
 \hat H\psi_n \equiv (\frac
 {\alpha_i}{i}\frac{\partial}{\partial z_i}+\beta
 m)\psi_n(\vec z) + \beta \int M (p_4=0, \vec z, \vec z')\psi_n(\vec
 z') d^3 \bar z'= \varepsilon_n \psi_n (\vec z)
 \ee
 where
 $$\alpha_i=\gamma_5\Sigma_i=\left(
 \begin{array}{ll}
 0&\sigma_i\\
 \sigma_i&0
 \end{array}
 \right);
$$
with eigenfunctions $\psi_n$ satisfying usual orthonormality
condition
$$
\int \psi^+_n(x)\psi_m( x) d^3 x=\delta_{nm},
$$
The Green's function $S$ can be expressed as
\be
S(p_4,\vec x, \vec y)=\sum_n\frac{\psi_n(\vec x)\psi^+_n(\vec
y)}{p_4\gamma_4-i\varepsilon_n\gamma_4}
\ee
Inserting (46) into (44) one is met with integrals of the type:
\be
\int^{\infty}_{-\infty}\frac{dp'_4}{2\pi}\frac{e^{-(p_4-p'_4)^2T^2_g}}
{(p'_4
\gamma_4-i\varepsilon_n\gamma_4)}=\frac{i}{2}\gamma_4
sign\varepsilon_n(1+0(p_4T_g,|\varepsilon_n|T_g)
\ee
Note, however, that the result depends on the boundary conditions.
If,
e.g., one imposes the causality--type boundary condition,
then one obtains
$$
\int\frac{dp'_4}{2\pi}\frac{e^{ip'_4h_4}}{\gamma_4(p'_4-i\varepsilon_n)}=
\left\{
\begin{array}{ll}
i\gamma_4e^{-\varepsilon h_4}\theta(\varepsilon_n),& h_4>0\\
-i\gamma_4\theta(-\varepsilon_n)e^{\varepsilon h_4},& h_4<0
\end{array}
\right.
\eqno{(47')}
$$

We are  thus led  to the following expression for $M$ in the string
limit
\be
M(p_4=0, \vec z, \vec w)= \sqrt{\pi} T_g[J_{ii}^{\mu\mu}(\vec z,\vec
w)\gamma_{\mu}\Lambda(\vec z, \vec w)
 \gamma_4\gamma_{\mu}-J^{ik}_{ik}(\vec z, \vec w)
\gamma_k\Lambda(\vec z, \vec w) \gamma_4\gamma_i]
\ee
where the definition is used
\be
\Lambda(\vec z, \vec w) = \sum_n\psi_n(\vec z) sign (\varepsilon_n)
\psi^+_n(\vec w)
\ee

Let us  disregard for the moment the possible appearance in $M$ of
the vector component (proportional to $\gamma_{\mu},\mu=1,2,3,4)$ and
concentrate on the scalar contribution only, since that is
responsible for CSB. Then one can look for solutions of the Dirac
equation (45) in the following form [17]
\be
\psi_n(\vec r)=\frac{1}{r}\left (
\begin{array}{l}
G_n(r)\Omega_{jlM}\\
iF_n(r)\Omega_{jl'M}
\end{array}
\right)
\ee
where $l'=2j-l$, and introducing the parameter
$\kappa(j,l)=(j+\frac{1}{2}) sign (j-l)$, and replacing $M$ by a local
operator (the generalization to the nonlocal case is straightforward
but cumbersome, the final result (53) is not changed in the nonlocal
case). We obtain a system of equations [17]
\be
\left\{
\begin{array}{l}
\frac{dG_n}{dr}+\frac{\kappa}{r}G_n-(\varepsilon_n+m+M(r))F_n=0\\
\frac{dF_n}{dr}+\frac{\kappa}{r}F_n-(\varepsilon_n-m-M(r))G_n=0\\
\end{array}
\right.
\ee
Eq.(51) possesses a symmetry $(\varepsilon_n, G_n,
F_n,\kappa)\leftrightarrow
(-\varepsilon_n,
F_n,G_n,-\kappa)$
which means that for any solution of the form (50) corresponding to
the eigenvalue $\varepsilon_n$, there is another solution of the form
\be
\psi_{-n}(r)=\frac{1}{r}\left (
\begin{array}{l}
F_n(r)\Omega_{jl'M}\\
iG_n(r)\Omega_{jlM}
\end{array}
\right )
\ee
corresponding to the eigenvalue $(-\varepsilon_n)$.

Therefore the difference, which enters (49) can computed in terms of
$F_n,G_n$ as follows
$$
\Lambda(\vec x,\vec y) =\frac{1}{xy}\beta
\sum_{njlM}[G_n(x)G^*_n(y)\Omega_{jlM}(\vec x)\Omega^*_{jlM}(\vec y)
-
$$
$$
-F_n(x)F_n^*(y)\Omega_{jl'M}(\vec x)\Omega_{jl'M}(\vec y)] +
\frac{i\gamma_5}{x y}
\sum_{njlM}\{F_n(x)G^*_n(y)\times
$$
\be
\times
\Omega_{jl'M}(\vec
x)\Omega^*_{jlM}(\vec y)
- G_n(x) F^*_n(y)\Omega_{jlm}(\vec x)\Omega^*_{jl'M}(\vec y)\}
\ee
The expression (53) serves to display the appearence of the
$\gamma_4$ factor in the main term, which as one can undestand from
(48), produces the scalar contribution for $M(\vec z, \vec w)$. For
quarks of heavy mass the sum (53) reduces to the $\delta$--function
term, which can be most easily seen in the simplified example, taking
$M$ to be constant. In this case one can write
\be
\Lambda(\vec x,\vec y) = \int \frac{d\vec p}{(2\pi)^3} e^{i\vec
p(\vec x - \vec y)}
\frac{\beta(m+M)+\vec \alpha\vec p}{\sqrt{\vec p^2+(m+M)^2}}
\ee
One can easily see from (54), that for $(m+M)\to \infty$ one has
\be
lim \Lambda_{m+M\to \infty}(\vec x,\vec y)=
\gamma_4\delta^{(3)}(\vec
x-\vec y)
\ee

We now turn to the $J_{ik}(\vec z,\vec z)$, (13), which can be
rewritten as
$$
J_{ik}^{44}(\vec z,\vec z)=z_iz_k\int^1_0\int^1_0 dt dt'D(\vec
z(t-t')),
$$
\be
J_{ik}^{ik}(\bar z,\bar z)=z_iz_k\int^1_0\int^1_0t dt t'dt'D(\vec
z(t-t')).
\ee
For the Gaussian form (17) one obtains at large $|\bar z|$
\be
J_{ik}^{44}(|\vec z|\to \infty ) = \frac{z_iz_k}{|\vec z|}
2T_g\sqrt{\pi}D(0),~~
J_{ik}^{ik}=\frac{1}{3} \frac{z_iz_k}{|\vec z|}
2T_g\sqrt{\pi}D(0)
\ee
Inserting this into (48) one gets for $M(\vec z,\vec z)$ at large
distances $|\vec z|$
\be
M(p_4=0,\vec z,\vec w) =\sigma |\vec z|(2-\frac{1-\gamma_4}{3})
\delta^{(3)}(\vec z-\vec w)
\ee
Thus one obtains seemingly a local Dirac equation for $\Psi_n$ and
$S$ with the mostly scalar kernel, containing linear confinement.

However, the kernel (58) does not correspond to the Dirac equation
for a quark of large mass $(m+M)$  (the latter property was used to
replace $\Lambda(\vec x, \vec y)$ by the $\delta$--function, as in
(55)).

To see this, one should keep in mind that the kernel (58) is the
limit $p_4T_g\to 0$ of the full kernel $M(p_4,\vec z, \vec w)$,
Eq.(44), while for the large mass $m$ the effective $p_4$ is arround
$m$ and therefore also large and the local Dirac equation obtains in
the limit, when one uses instead of equations (43), (44), the set
(15), (16), where in the lowest approximation in $M/m$ one inserts in
(15) the free   propagator for the heavy particle $m$, Eq. (34).
Details of this derivation and results are given in Appendix 5.

One can see that the resulting Dirac equation has the effective
kernel (38), with the  expected coefficient $\sigma |\vec z|$, in
contrast to (58). Therefore we shall see that the expression (58) is
valid only for light quarks, $m T_g\ll 1$,  and for the case when
$\Lambda (\vec z,\vec w)$ reduces to the $\delta$ -- function term.
To see when it is possible one can use the quasiclassical
approximation to calculate $ \Lambda (\vec z, \vec w)$ and we do it
in the next chapter, while we conclude this chapter with the analysis
of the $q\bar Q$ spectrum, resulting from the spectrum of the
Hamiltonian (45).

There is an important point, which one should have in mind concerning
the Dirac equation. As we have discussed above in this section, the
Dirac equation with the  scalar interaction produces the spectrum
symmetric under reflection $\varepsilon_n\to -\varepsilon_n$,
 and it is clear that negative energy states are not present  in
 the spectrum of the heavy--light mesons in reality.  To resolve
 this paradox,  let  us come back to our Green's function
 \be
 S_{q\bar Q}= (\Gamma  S (x_4-y_4, \vec z,\vec y) \Gamma S_{\bar
 Q}(x_4-y_4))
 \ee
 The masses of the $(q\bar Q)$ system are obtained from the
 asymptotics of $S_{q\bar Q}$  at
 $ T\equiv x_4-y_4\to \infty,$ i.e. $S_{q\bar Q}\sim exp (-E_n T)$
 one should put in (59) $S_{\bar Q}$  as
 \be
 S_{\bar Q}(h)= i[\frac{1-\gamma_4}{2} \theta (h) e^{-M_Qh}+
 \frac{1+\gamma_4}{2}\theta(-h)e^{M_Q}h]
 \ee
  and considering positive $T=x_4-y_4$, one can express $G_{q\bar
  Q}$ as
  $$
    G_{q\bar Q}=\sum_n tr\{\Gamma\gamma_4\int
  \frac{|n><n|}{p_4-i\varepsilon_n}
  e^{ip_4T-M_QT}\frac{dp_4}{2\pi}\frac{(1-\gamma_4)}{2}\Gamma\}
  =
  $$
  \be
  =\sum_n tr (\Gamma \gamma_4|n>\theta (\varepsilon_n)<n|
  e^{-(M_Q+|\varepsilon_n|)T}
  \frac{(1-\gamma_4)}{2}\Gamma)
  \ee
  Hence one can see that only positive values of  $\varepsilon_n$
  contribute to the mass of the $q\bar Q$ system, namely
  $$
  M_{q\bar Q}=M_Q+|\varepsilon_n|,
  $$
  Using (48), (49) and (45), one can write a nonlinear equation for
  eigenfunctions $\psi_n$, namely
  $$
  (-i\alpha_i \frac{\partial}{\partial z_i}+\beta m)\psi_n (\vec z)+
  \int \beta [\tilde J^{\mu}_{ik} (\vec z, \vec w) \gamma_{\mu}\sum_k
  \psi_k (\vec z) sign (\varepsilon_k) \psi^+_k(\vec w)\times
  $$
  $$
  \times \gamma_4\gamma_{\mu}-\tilde J^{ik}_{ij}(\vec z, \vec w)
  \gamma_j\sum_k \psi_k(\vec z) sign (\varepsilon_k) \psi^+_k(\vec w)
  \gamma_4 \gamma_i]\psi_n(\vec w) d^3 w=
  $$
  \be
  =
  \varepsilon_n\psi_n (\vec z)
  \ee
  where we have  defined
  \be
  \tilde J_{ik}^{\mu(ik)} =\sqrt{\pi} T_g J_{ik}^{\mu(ik)}
  \ee
  and $J^{\mu}_{ik}$ is given in (13).

  \section{Quasiclassical solution of the selfconsistent equation}

  In this chapter the quasiclassical analysis of equations (62) will
  be given. The nonlinear part of the kernel $M$, Eq. (48), is
  $\Lambda(\vec x, \vec y)$ which can be expressed through solutions
  $\psi_n$ as in (53). Our primary task is now to calculate
  $\Lambda(\vec x, \vec y)$ in some reasonable approximation, having
  in mind to improve it at later step.

   Therefore we start  calculating $\Lambda(\vec x, \vec y)$ in
   the form (53) using  for $G_n$, $F_n$  quasiclassical solutions of
   the local Dirac equation with the scalar potential $U=\sigma
   r$.  As one will see the resulting $\Lambda(\vec x, \vec y)$
   is the quasilocal object, tending to the $\delta(\vec x-\vec
   y)$ at large $|\vec x|$. Being inserted in (62) or (48),
   $\Lambda(\vec x, \vec y)$ would indeed generate a local kernel
   $M(\vec z, \vec w)$, producing the local (at large $r$) Dirac
   equation with $U(r)$ tending to $\sigma r$ at large distances.
   Thus one justifies a posteriori the initial choice of $G_n,F_n$
   at  least for large $r$.

   Coming back to calculating $\Lambda(\vec x,\vec y)$ for $U=\sigma
   r$, we follow the method of [18] and write the Dirac equation as
   follows
   \be
   \frac{d}{dr}\psi=\frac{1}{\hbar} D\psi; \psi = \left(
   \begin{array} {l}
   G\\
   F
   \end{array}
   \right),
   ~D=\left(
   \begin{array}{l}
   -\frac{\kappa}{r},~~ m+U(r)+\varepsilon-V\\
   m+U-\varepsilon+V,~~ \frac{\kappa}{r}
   \end{array}
   \right)
   \ee
   We have kept the vector potential $V$ in (64) to make our
   consideration as general as possible.

   To do the quasiclassical expansion, one writes [18]
   $$
   \psi=\varphi exp \int^r y(r) dr,~~ \varphi=
   \sum^{\infty}_{n=0}\hbar^n\varphi^{(n)},
   $$
   \be
   y(r)=\frac{1}{\hbar}y_{-1}(r)+y_0(r)+\hbar y_1(r)+...
   \ee
   and obtains the system of equations
   $$
   (D-y_{-1})\varphi^{(0)}=0
   $$
   $$
   (D-y_{-1})\varphi^{(1)}=\frac{d}{dr}\varphi^{(0)}+y_0\varphi^{(0)},
      $$
   \be
   (D-y_{-1})\varphi^{(n+1)}=\frac{d}{dr}\varphi^{(n)}+\sum^n_{k=0}
   y_{n-k}\varphi^{(k)}
   \ee
   In what follows we keep in the lowest approximation
   $y_{-1}(r),y_0(r)$ and $\varphi^{(0)}$, and obtain
   \be
   y_{-1}\equiv \lambda_i=\pm
   \sqrt{(m+U)^2+\frac{\kappa^2}{r^2}-(\varepsilon-V)^2}\equiv \pm q
   \ee
   \be
   y_0^{(i)}(r)=-\frac{\lambda'_i}{2\lambda_i}+
   \frac{\kappa}{2\lambda_ir^2}-
   \frac{U'-V'}{2(m+U+\varepsilon -V)}+
   \frac{\kappa}{2\lambda_ir}\frac{(U'-V')}{(m+U+\varepsilon -V)},
   \ee
   \be
   \varphi_i^{(0)}=A\left(
   \begin{array} {l}
   m+U+\varepsilon-V\\
   \lambda_i+\frac{\kappa}{r}
   \end{array}
   \right)
   =A'\left(
   \begin{array}{l}
   \lambda_i-\frac{\kappa}{r}\\
   m+U-\varepsilon+V
   \end{array}
   \right),
   \ee
   and
   \be
   \psi^{(i)}=\varphi_i^{(0)} exp \int^r(y^{(i)}_{-1}(r')
   dr'+y_0^{(i)}(r') dr')
   \ee
   where the index $i=+,-$ refers to two possible solutions of the
   matrix equation $(D-y_{-1})\varphi^{(0)}=0$.

From (67) one can find three different regions on the line $0 \leq r
< \infty$ ( for $U>|V|$). Taking for simplicity $m=0,V=0$, one has
from $q=0$,
\be
\varepsilon^2 = \sigma^2 r^2+\frac{\kappa^2}{r^2},~~
r^2_{\pm}=\frac{\varepsilon^2\pm
\sqrt{\varepsilon^4-4\sigma^2\kappa^2}}{2\sigma^2}
\ee
Thus for $\varepsilon^2 >2\sigma \kappa$ ( and this holds for all
levels, see below) one obtains two turning points $r_{\pm}$ in (71).
Between these points $\lambda_i$ is  imaginary and this is the
classically allowed region with the momentum $p(r)$
\be
p(r) = \sqrt{(\varepsilon-V)^2-\frac{\kappa^2}{r^2}-(m+U)^2}
\ee
The quasiclassical solution in this region can be written in analogy
with the corresponding solution in [18, first entry Eq. (1.12)]
\be
G=C_1\sqrt{\frac{m+U+\varepsilon-V}{p(r)}}sin \theta_1; F=C_1 sgn
\kappa\sqrt{\frac{\varepsilon-V-m-U}{p(r)}} sin \theta_2
\ee
where we used the notation similar to that of [18]
\be
\theta_1(r)=\int^r_{r_-}(p+\frac{\kappa w}{pr'})dr'
+\frac{\pi}{4},\theta_2(r) =
\int^r_{r_-}(p+\frac{\kappa \tilde w}{pr'}) dr'+\frac{\pi}{4}
\ee
\be
w(r)=-\frac{1}{2r}-\frac{1}{2}\frac{U'-V'}{m+U+\varepsilon -V}
\ee
\be
\tilde w(r)=\frac{1}{2r}-\frac{1}{2}\frac{U'-V'}{m+U+\varepsilon -V}
\ee

Here prime denotes the derivative in $r$;  putting $U\equiv 0$ one
recovers formulas(1.12-1.14) of [18].

 As a next step we turn to the quasiclassical
 determination of energy eigenwalues $\varepsilon_n$.
 Using the Bohr--Sommerfeld equation
 \be
 \int^{r_+}_{r_-}p(r)dr=\pi(n+\frac{1}{2}),~~n=0,1,2...,
 \ee
 one obtains for $m=0,V=0$, with $p(r)$ from (72)
 \be
 \varepsilon^2_n=4\sigma(n+\frac{|\kappa|+1}{2})=
 2\sigma (2n+j+\frac{3}{2})
 \ee

 Comparison  of the numbers obtained from (78) with exact calculation
 of Dirac equation in [17] reveals that (78) is a very poor
 approximation  for the real spectrum,  and, moreover, it is
 qualitatively incorrect, since $\varepsilon_n$ in (78) does not
 depend on the sign of $\kappa$, i.e. on the spin-orbit interaction
 of the light quark. Happily this problem was already treated in [18]
 for Coulomb interaction,
 and the authors have proved that it is legitimate to take  into
 account spin--orbit interaction quasiclassically; they suggested
 another eigenvalue condition instead of the Bohr--Sommerfeld, which
 can be deduced from the solutions (73), namely
 \be
 \int^{r_+}_{r_-}(p+\frac{\kappa w}{pr}) dr =\pi(n+\frac{1}{2}),
 n=0,1,2,...
 \ee
 where $w$ is defined in (75).
 Doing the integrals approximately in (79), one obtains the following
 spectrum  instead of (78).
 \be
 \varepsilon^2_n=4\sigma(n+\frac{1}{2}+\frac{1}{2}j+\frac{1+sgn\kappa}{4}
 -\frac{\kappa\sigma}{2\pi\varepsilon^2_n}(0.38+ln
 \frac{\varepsilon^2_n}{\sigma|\kappa|}))+0
 \left((\frac{\kappa\sigma}{\varepsilon^2_n})^2\right)
 \ee
 One can visualize the appearence of the last two terms, depending on
 the sign of $\kappa$, which produce the spin-orbit splitting of the
 levels. The form (80) is exact for small
 $(\frac{\kappa\sigma}{\varepsilon^2_n})^2$, and for lowest levels
 the accuracy can be tested by comparison  with the exact solution of
 Dirac equation [17]; two sets of 6 lowest levels coincide within 5\%,
 whereas for (78) the accuracy could be as bad as $25\div 30$\%. Note
 however, that for high excited levels, where $n$ and /or $|\kappa|$
 are large, the correction terms in (80) are constant, and therefore
 relatively unimportant as compared with growing $n$ and $j$.
 Therefore in what follows in summation over
 high excited states we shall use the simplified form (78) instead of
 the corrected one, Eq.(80).

 We turn now to the quasiclassical calculation of $\Lambda(\vec
 x,\vec y)$, Eq.(53), and to this end we represent it in the
 following form
 \be
 \Lambda(\vec x,\vec y)=\beta\Lambda_1(\vec x, \vec y)
 +i\gamma_5\Lambda_2(\vec x, \vec y)
 \ee
 where, e.g., $\Lambda_1$ is a 2x2 matrix, $\mu,\mu'=\pm \frac{1}{2}$
 $$
 \Lambda^{\mu\mu'}_1(\vec x, \vec y)=\frac{1}{xy}\sum_{njlM}\{
 G_n(x)G_n^*(y)\Omega^{\mu}_{jlM} (\vec x)\Omega_{jlM}^{*\mu'} (\vec
 y)-
 $$
 \be
 -F_n(x)F_n^*(y)\Omega^{\mu}_{jl'M}(\vec x)\Omega^{*\mu'}_{jl'M}(\vec
 y)\},
 \ee
 and the sum is over positive energy states only; $G_n,F_n$ depend
 not only on $n$, but also on $j,l$ but independent on $M$.

 We shall be interested mostly in the large values of $n,l$ in the
 sum (82), since those terms will form a tempered $\delta$--function
 of the type $\delta^{(3)}(\vec x-\vec y)$. Hence one can neglect in
 $\theta_1,\theta_2$, Eq. (74) the constant terms $\frac{\kappa
 w}{pr}, \frac{\kappa\tilde w}{pr}$ as compared to the growing
 contribution $\int pdr'$. Now in $p(r)$, (72), one can replace
 $\kappa^2$ by $l^2$ at large $l$, and as a result both $G_n$ and
 $F_n$ in (82) do not depend on $j$ (with $l$ fixed) and one can sum
 up over $j,M$
 $$
 \sum_{jM}\Omega^{\mu}_{jlM}(\vec x)\Omega^{\mu'*}_{jlM}(\vec y)=
 \delta_{\mu\mu'}\sum_M Y_{l,M-\mu}(\vec x)Y^*_{l,M-\mu}(\vec y)=
 $$
 \be
 =\delta_{\mu\mu'}\frac{2l+1}{4\pi}P_l(cos\theta_{xy})
 \ee

 Hence $\Lambda^{\mu\mu'}_1$ is diagonal matrix
 $\Lambda^{\mu\mu'}_1=\delta_{\mu\mu'}\Lambda_1$
 with $\Lambda_1$ equal to
 \be
 \Lambda_1(\bar x, \vec
 y)=\sum_{n,l}C^2_1(n,l)\frac{2l+1}{4\pi}P_l(cos \theta_{xy})\{[
 N_1(x)N_1(y)
 -N_2(x)N_2(y)]\sin \bar \theta (x) sin \bar \theta (y)\}
 \ee
 where we have defined
 \be
 N_1(x)=\sqrt{\frac{\varepsilon_n+m+U(x)-V(x)}{p(x)}};
 N_2(x)=\sqrt{\frac{\varepsilon_n-m-U(x)-V(x)}{p(x)}}
 \ee
 and
 $
 \bar \theta (x)=\int^x_{r_-}pdr.$

 The normalization constant $C_1(n,l)$ entering (84), can be
 estimated as
 \be
 C_1^2\int^{r_+}_{r_-}
 dr \{\frac{m+U+\varepsilon-V}{p(r)} sin^2 \theta_1+
 \frac{\varepsilon-V-m-U}{p(r)} sin^2 \theta_2\}=1
 \ee

For large $l,n$ the functions $ sin \theta_1(r), sin \theta_2(r)$ are
 fast oscillating and hence one can replace
 \be
 sin^2 \theta_1\approx sin^2 \theta_2\approx \frac{1}{2}
 \ee
 and one obtains for $V\equiv 0$,
 \be
 C^{-2}_1 = \varepsilon \int^{r+}_{r-}\frac{dr}{p(r)}=
 \frac{\varepsilon\pi}{2\sigma};~~
 C^{2}_1(n,l) =
 \frac{2\sigma}{\varepsilon_n\pi}
\ee
Consider now in $\Lambda_1(\bar x,\bar y)$ both $x,y$ in the region
\be
r_-\ll x , y <r_+,~~ |x-y|\ll x,y
\ee
Then one has
\be
N_1(x)N_1(y)-N_2(x)N_2(y)\approx \frac{2(m+U(x))}{p(x)}
\ee
and
$$
sin \bar \theta (x) sin \bar \theta (y) =\frac{1}{2} [cos (\bar
\theta (x) -\bar \theta(y))- cos (\bar \theta(x)+\bar
\theta(y))]\approx
$$
\be
\approx \frac{1}{2} cos \int ^x_y\sqrt{\varepsilon^2-\sigma^2
r^2-\frac{\kappa^2}{r^2}}dr +...
\ee
where dots stand for a fast oscillating term, which will not
contribute to the final answer.
Hence one obtains for $\Lambda_1$,
\be
\Lambda_1(\vec x, \vec y)=\frac{1}{xy}\sum_{l,n}
\frac{2\sigma}{\pi\varepsilon_n}
\frac{2l+1}{4\pi}P_l(cos\theta_{xy}) \frac{m+U(x)}{p(x)} cos \int^x_y
p(r)dr
\ee
One can compute the integral under the cosine
\be
I\equiv
\int^x_yp(r)dr=\frac{1}{2}(p(x)x-p(y)y)-
\frac{\varepsilon^2}{4\sigma}(\alpha
(x)-\alpha(y))-\frac{|\kappa|}{2}(\beta(x)-\beta(y))
\ee
where $p(x) $ is given in (72) with $V\equiv 0$, and
\be
\alpha(x)=arc\sin\frac{\varepsilon^2-2\sigma^2x^2}
{\sqrt{\varepsilon^4-4\sigma^2\kappa^2}},
\beta(x) =arc\sin
\frac{\varepsilon^2-\frac{2\kappa^2}{x^2}}
{\sqrt{\varepsilon^4-4\sigma^2\kappa^2}}
\ee
In what follows we replace the sum over $n$ in (92) by integration
 over $\varepsilon$, with $l$ (or $|\kappa|)$ fixed and neglect in
 (93) terms containing $\kappa$,
 which are small at large $n$.

  We also introduce instead of $\varepsilon$ the variable $\tau$ as
  follows
  \be
  \varepsilon=\sigma x \tau; \tau\geq 1; p(x)=\sigma x\sqrt{\tau^2-1}
  \ee
  From (80) one has asymptotically
  \be
  \sum_n\to \int \frac{\varepsilon d\varepsilon}{2\sigma}
  \ee
  Expanding $\alpha(x,y)$ in (93) around $x=y$  one has
  \be
  I(x,y) =p(x)(x-y)+0((x-y)^2)=\sigma x(x-y)\sqrt{\tau^2-1}
  \ee
  Finally one obtains in (92)
  $$
  \Lambda_1(x,y)\cong
  \frac{\sigma}{x}\sum_l\int^{\infty}_1\frac{d\tau}{\pi}
  \frac{2l+1}{4\pi}P_l(cos
  \theta_{xy})\frac{cos(a\sqrt{\tau^2-1})}{\sqrt{\tau^2-1}}=
  $$
  \be
  =\frac{\sigma}{\pi
  x}\sum_l\frac{2l+1}{4\pi}P_l(cos\theta_{xy})K_0(a)
  \ee
  where
  \be
  a=\sigma x|x-y|
  \ee
   and $K_0$ is the McDonald function.

    Now one can perform the sum over $l$ in (98) to get
    \be
    \sum(2l+1)P_l(cos\theta)=4\delta(1-cos\theta)
    \ee
    Thus one obtains for large $x,y$ and $|x-y|\ll x,y$
 the result, which we write in the form symmetric in $x,y$
     \be
     \Lambda_1(\vec
     x,\vec y)_{x\sim y}\cong\frac{\sigma}{\pi^2\sqrt{xy}}
     K_0 (\sigma\sqrt{xy}|x-y|)\delta(1-cos\theta_{xy})
      \ee
       One can
     test that $\Lambda_1(\vec x,\vec y)$ is a smeared $\delta$ --
    function, normalized to 1, with the smearing radius
     \be
r_0\approx (\sigma\sqrt{xy})^{-1}\approx (\sigma
      x)^{-1}\ll min (\frac{1}{\sqrt{\sigma}},x)
      \ee
      The normalization can be checked doing the integral:
\be
\int d^3y\Lambda_1(x,y)=\int dcos \theta \delta(1-cos \theta)
      d\varphi_y\frac{x^2dy}{\pi^2 x} K_0(a)
      \ee
writing
\be
y=x\pm\frac{a}{\sigma x},~~dy=\pm\frac{da}{\sigma x}
\ee
and taking into account that
\be
\int d~ cos\theta \delta (1-cos\theta)=\frac{1}{2}
\ee
\be
\int^{\infty}_0 da K_0(a)=\frac{\pi}{2}
\ee
one obtains
\be
\int d^3y\Lambda_1(\vec x,\vec y)=1
\ee

Thus $\Lambda_1(\vec x, \vec y) $ is a normalized smeared
$\delta$--function, which is "focusing" the nonlocal interaction
kernel $M(\vec x, \vec y)$, Eq. (48), at least at large distances
$x,y$ into a quasilocal kernel, linearly growing at large $x\approx
y$.

At the same time another part of $\Lambda(\vec x, \vec y)$, namely
$\Lambda_2(\vec x, \vec y),$ (see Eqs. (53), (81)) can be written as
$$
\Lambda^{\mu\mu'}_2(\vec x,\vec y)=\frac{1}{xy}\sum_{njlM}\{
F_n(x)G_n^*(y)\Omega^{\mu}_{jl'M}(\vec x)\Omega^{*\mu'}_{jlM}(\vec
y)-
$$
\be
-G_n(x)F_n^*(y)\Omega^{\mu}_{jlM}(\bar x)\Omega^{*\mu'}_{jl'M}(\vec
y)\}
\ee
Doing the same procedure with $\Lambda_2$ as was done with
$\Lambda_1$, and to this end using the relation
\be
\Omega^{\mu}_{jl'M}(\vec x) =-(\vec \sigma\vec
n_x)_{\mu\nu}\Omega^{\nu}_{jlM}(\vec x)
\ee
one obtains
$$
\Lambda_2^{\mu\mu'}(\vec x,\vec y)=\frac{-1}{xy}\sum_{nl}C_1^2(n,l)\{
N_2(x)N_1(y)sin \theta_2(x)sin \theta_1(y)(\vec \sigma \vec
n_x)_{\mu\mu'}
-
$$
\be
-N_1(x)N_2(y) sin \theta_1(x) sin \theta_2(y) (\vec \sigma\vec
n_y)_{\mu\mu'}\} \frac{2l+1}{4\pi}P_l(cos \theta)
\ee
One can see from (110) that $\Lambda_2(\vec x, \vec y)$ is odd with
respect to exchange $\vec x\leftrightarrow\vec y$ and therefore
vanishes at $\vec x =\vec y$. Hence $\Lambda_2(\vec x, \vec y)$  has
no local limit  and gives no contribution to the long distance linear
interaction $\Lambda(\vec  x, \vec x)$.

We shall not be interested in $\Lambda_2$ for the rest of the paper,
however it may contribute to the finite  range nonlocal part of the
resulting kernel $M(\vec x, \vec y)$.

\section{Calculation of the chiral condensate}

Our method allows to calculate the chiral condensate -- the
characteristics, which does not depend on the presence of the static
quark.
To this end we consider the euclidean condensate, which due to the
definition (14) is
\be
<\psi^+(x)\psi(x)>= N_c tr S(x,x)
\ee
 At this point it is important that we take in $S(x,y),~x=y=0$, thus
 putting the initial and final point on the trajectory of the static
 quark, and not at the point $\vec x=\vec y$ far from origin. The
 reason is that with our choice of the gauge, (5),(6) and in the
 Gaussian approximation (12) the string from the light quark extends
 from the  point $\vec x, \vec y$  necesarily to the $\vec x =0$ and
 therefore the world sheet of the string is not the minimal surface,
 but rather a cone with the vertex at the point $\vec x = 0$.
The choice $S(0\vec x=0,\vec y=0)$ makes the area inside the
characteristic trajectory of $S$ as minimal as possible, thus
greately diminishing the contribution of non-Gaussian correlators.

The latter collectively are responsible for giving the minimal area
contribution for any choice of the point $z_0$ in the Fock--Schwinger
gauge condition.

To proceed one can use our  definitions (46), (47), (49)  and (81) to
write
$$
\frac{1}{N_c}<\psi^+(0)\psi(0)>= tr S(h_4=0,\vec 0,\vec 0)=
$$
\be
\int\frac{dp_4}{2\pi} tr S(p_4,\vec 0, \vec 0)=
{i}\Lambda^{\mu\mu}_1(\vec 0,\vec 0)
\ee
Using (82), one can rewrite (112) as
\be
\frac{1}{N_c}<\psi^+\psi>=2i\frac{1}{4\pi}\sum_{njl}\{
(\frac{G_n(x)}{x})^2_{x=0} -(\frac{F_n(x)}{x})^2_{x=0}
\}
\ee
The analysis of the behaviour of $G_n(x),F_n(x)$ as solutions of the
Dirac equation with the linear potential at small $x$ was performed
in [17], where the following properties were found
\be
G=Ax^{|\kappa|}+..., F=Bx^{|\kappa|},
\ee
Thus in (113) the contribution comes only from the states
$|\kappa|=1$, i.e. $j={1/2}$ and $l$ or $l'=0$.

In the first case
\be
1)
\kappa=-1, l=0; A\neq 0, B=0
\ee
In the second case
\be
2) \kappa =1, l=1; l'=0; A=0, B\neq 0
\ee

Denoting
\be
A(n,\kappa=-1)=A^-_n, B(n, \kappa=+1)=B^+_n
\ee
and heaving in mind the relation between Minkowskian and Euclidean
quark condensate
\be
<\psi\bar \psi>_M=i<\psi \psi^+>_E
\ee
one obtains the general expression
\be
<\psi\bar\psi>_M=-\frac{N_c}{2\pi}\sum^{\infty}_{n=0}[(A_n^-)^2-(B_n^+)^2]
\ee
Our next step is the calculation of $A_n^-,B_n^+$ using the
quasiclassical method. The standard matching condition connects the
wave function below and above the turning point $r=r_-$:
\footnote{
Recently an important correction was found [19] to the  standard
nonrrelativistic matching condition, which is not taken into account
below, since at large $n$ it is small.}
 \be
  \frac{C_1}{\sqrt{p}}cos
(\int^r_{r_-}(p+\frac{\kappa w}{pr'}) dr'-\frac{\pi}{4})\to
\frac{C_1}{2\sqrt{|p|}}\exp (-|\int^r_{r_-}(p+ \frac{\kappa w}{pr'})
dr'|)
 \ee
  Denoting
   \be
    \bar\theta_1(r)=\int^r_{r_-}(p+ \frac{\kappa
w}{pr'}) dr', \bar\theta_2(r)=\int^r_{r_-}(p+ \frac{\kappa \tilde
w}{pr'}) dr'
 \ee
  one has the following behaviour of $\bar
\theta_1(r)$ at small $r$:
 \be
 \bar\theta_1(r)=
i(|\kappa|+\frac{1}{2} sign
\kappa)ln\frac{r}{2r_-}+is_1
\ee
where $s_1$ is finite for $r\to 0$.

Similarly for $\bar \theta_2(r)$ one obtains
\be
\bar\theta_2(r)=
i(|\kappa|+\frac{1}{2} sign
\kappa)ln\frac{r}{2r_-}+is_2
\ee
From (120),(122) and (123) the  the $A^-_n$ and $B^+_n$ are
\be
A^-_n=\frac{c^-_n\sqrt{\varepsilon^-_n}}{2\sqrt{2r_-}}e^{s_1},
B^+_n=\frac{c^+_n\sqrt{\varepsilon^+_n}}{2\sqrt{2r_-}}e^{s_2}
\ee
where the following notations are used
\be
\left\{
\begin{array}{l}
\varepsilon^+_n\equiv \varepsilon(n,\kappa=+1), \varepsilon^-_n=
\varepsilon(n,\kappa=-1)\\
c^{\pm}_n=C_1(n,\kappa=\pm 1)
\end{array}
\right.
\ee
To the lowest order in $1/n$ expansion one has
$$
s_1=|\kappa|-\frac{\kappa}{2\varepsilon
r_+}\frac{\pi}{2}=1+\frac{\pi}{4\varepsilon r_+}+0(\frac{1}{n^2})
$$
\be
s_2=|\kappa|-\frac{\kappa}{2\varepsilon
r_+}\frac{\pi}{2}=1-\frac{\pi}{4\varepsilon r_+}+0(\frac{1}{n^2})
\ee

Since for large $n$ one has
\be
(\varepsilon^{\pm}_n)^2=4\sigma(n+1\pm\frac{1}{4}+0(\frac{1}{n}))
\ee
and due to (88)
$C^{\pm}_n=(\frac{2\sigma}{\varepsilon^{\pm}_n\pi})^{1/2},
$
and
$r_-=\frac{1}{\varepsilon_n}(1+0(\frac{1}{n^2}))$, the difference
$(A_n^-)^2-(B_n^+)^2$
vanishes at large $n$ in the leading order, namely one obtains
$$
(A_n^-)^2-(B_n^+)^2=\frac{2\sigma
e^2}{\pi}\{\varepsilon^-_n-\varepsilon^+_n+
\frac{\pi\sigma}{2\varepsilon^-_n}+\frac{\pi\sigma}{2\varepsilon^+_n}\}
=
$$
\be
=\frac{e^2\sigma^{3/2}}{\pi\sqrt{n+1}}(\pi-1)
\ee
In (128) the factor in brackets occurs from subleading terms in
$\varepsilon^{\pm}_n$ (the term (-1)) and from the corrections
$s_1,s_2$ (the term $\pi$). However the corrections of the order of
$1/n$ in $c^{\pm}_n$ are not taken into account in (128), moreover,
the next order quasiclassical expansion terms, i.e. $y_1(r)$ and
$\varphi^{(1)}$ may contribute  to the $1/n$ correction to the
coefficients $A_n^-,B^+_n$, and the problem of computing the
difference (128) becomes rather complicated.

With all corrections included, the difference $(A_n^-)^2-(B_n^+)^2$
is of the order $0(\frac{1}{\sqrt{n}})$ ( modulo unexpected
cancellations), and the series (119) diverges. Therefore we must try
to approach the problem of computation of $S(0,0)\sim <\psi\bar
\psi>$ from another side. To this end we first consider the problem
of computing $\psi_n(0)$ for the nonrelativistic Schroedinger
equation with the local potential $\tilde U(r)$. Following [20], one
has
\be
|\psi_n(0)|^2=\frac{\mu}{2\pi}<\tilde
U'(r)>=\frac{\mu}{2\pi}\frac{\int d^3 r\psi^*_n\tilde
U'(r)\psi_n}{\int d^3 r|\psi_n|^2}
\ee
In particular, for the linear potential $\tilde U(r)=\sigma r,
|\psi_n(0)|^2=\frac{\mu\sigma}{2\pi}$ and does not depend on $n$. For
the quadratic potential one has $|\psi_n(0)|^2\sim <r>_{nn}\sim
\sqrt{n}$.

As one can see in (128), this case  is similar to ours, i.e. the
effective Schroedinger potential for the Dirac equation with linear
interaction is quadratic. This fact is well known in the
quasiclassical approximation [17,18], indeed one can write the equivalent
Schroedinger potential with the energy $\tilde E$, and effective
potential $\tilde U$ of the following form [17]
$$
\tilde E=\frac{1}{2}\varepsilon^2,~ \tilde U=\frac{1}{2}
U^2+\frac{\kappa^2}{r^2},~
p^2(r)=2(\tilde E-\tilde U),
$$
\be
(\frac{p^2(r)}{2}+\tilde U)\psi(r)=\tilde E\psi(r), \mu=1
\ee
Taking $U=\sigma r$, one  immediately obtains the leading term for
$(A^-_n)^2,(B^+_n)^2$ for large $n$, proportional to
$\sigma^{3/2}\sqrt{n}$. Thus far the effective nonrelativistic theory
with a local potential confirm our result (128) and  produces the
diverging sum as in (119).

Let us now consider the nonlocal effective  potential, $\tilde
U(r,r') $. The straightforward calculation similar to (129) yields in
this case
\be
|\psi_n(0)|^2=\frac{\mu}{\pi}\int^{\infty}_0 dr \int^{\infty}_0dr'
y^*_n(r)(\frac{d}{dr}\tilde U(r,r'))y_n(r')
\ee
where $y_n(r)$ is the radial part of $\psi_n(\vec r),$ normalized as
$\int^{\infty}_0y^2_n(r) dr=1$.
To understand to role of the nonlocality, let us represent $\tilde U$
as
 follows
 \be
 \tilde U(r,r')=(v(r)+v(r'))
 \frac{e^{-\frac{(r-r')^2}{a^2}} }{2
 a\sqrt{\pi}}
 \ee
 so that in the limit $a\to 0 $ one has
 \be
 \int^{\infty}_{-\infty} d(r-r') \tilde U(r,r') =v(r)
 \ee
 For the case when $v(r)$ corresponds to the equivalent effective
 potential, i.e.
 $v(r)=\frac{1}{2}\sigma^2 r^2, \mu=1$, see (130), we obtain
 \be
 |\psi_n(0)|^2=
 \frac{\sigma^2}{\pi}
 \int^{\infty}_0\int dr dr'
 y^*_n(r)
 \frac{r}{2a\sqrt{\pi}}
 e^{
 -\frac{(r-r')^2}{a^2}
 }y_n(r')
 \ee

 For the  chosen above potential $v(r)$ the wave function $y_n(r)$ is
 \be
 y_n(r)=\frac{C_n}{\sqrt{p(r)}}cos [
 (2n+1)(\frac{r}{r_+}\sqrt{1-(\frac{r}{r_+})^2}+arcsm\frac{r}{r_+})
 -\frac{\pi}{4}]
 \ee
 where
 $r^2_+=\frac{4}{\sigma}(n+1/2), C_n^2\approx \frac{4\sigma}{\pi}$,
 and
 \be
 p(r)=\sqrt{\varepsilon^2_n-\sigma^2r^2-\frac{\kappa^2}{r^2}}\approx
 \sqrt{\varepsilon^2_n-\sigma^2 r^2}
 \ee
 For $a\ll r_+$ and large $n$ one can approximate the integral (135)
 as follows
 \be
 |\psi_n(0)|^2=C^2_n\frac{\sigma^2}{\pi}\int\frac{d(\frac{r+r'}{2})r}{\sqrt
 {p(r)p(r')}}\int\frac{d(r-r')}{2a\sqrt{\pi}} cos
 [\frac{4n}{r_+}(r-r')\varphi(r,r')]exp (-\frac{(r-r')^2}{a^2})
 \ee
 where $\varphi(r,r')=1-\frac{r^2+rr'+r'^{2}
 }{6r^2_+}+0((\frac{r}{r_+})^3)$.
 The integral over $d(r-r')$ yields the cut-off factor
 \be
 \frac{1}{a\sqrt{\pi}}\int^{\infty}_{-\infty}d(r-r')cos
 (\frac{4n}{r_+}(r-r')\varphi) exp(-\frac{(r-r')^2}{a^2})=
 e^{-\frac{4n^2a^2\varphi^2}{r_+^2}}
 \ee
 and one has
 \be
 |\psi_n(0)|^2=\frac{\sigma\varepsilon_n}{\pi} exp
 (-\frac{4n^2a^2}{r_+^2}\bar \varphi^2)
 \ee
 where we have defined the effective value of $\bar\varphi^2$ of the
 order of one.

 Let us compare (139) with the square of $A^-_n, B_n^+$ from (124).
 The first factor in (139) coincides with that of (124) up to  a
 factor $(\frac{e^{2s}}{4})$, while the cut-off exponential factor is
 new and originates from the nonlocality of the interaction (132). It
 essentially cuts off the sum of (119) at the $n\geq n_{max}$,
 \be
 n_{max}\approx \frac{1}{\sigma a^2\bar\varphi^2}\sim \frac{1}{\sigma
 a^2}
 \ee
 We  turn now back to the calculation of the chiral condensate, Eq.
 (119), with the $A_n^2-B^2_n$ given by (128). The last expression is
obtained without corrections $0(1/n)$ to the normalization constants
$C^\pm_n$, which may change the numerical coefficient in (128), but
cannot change the $n^{-1/2}$ behaviour, leading to the divergence of
the sum (119).

We have noticed above, Eq. (139),(140), that nonlocality of the
effective potential, i.e. $M(p_4,\vec z, \vec w)$ causes the cut-off
of the sum (119) in $n$, and now one must look more closely at the
origin of this nonlocality.

To begin with we should remember, that the expression (82) for
$\Lambda_1(\vec x, \vec y)$ and as a result the sums in (113) and
(119) are obtained in the limit $T_g\to 0$,  when according to (44)
$\Lambda_1$ and $M(p_4,\vec z,\vec w)$ do not depend on $p_4$, so
that we have put $p_4=0$ in $M(p_4,\vec z, \vec w)$. Let us now take
the finite $T_g$ into account and compute $\Lambda_1(\vec x, \vec y)$
anew. To this end one can use the results of the Appendix 4, and for
simplicity we again put there $p_4=0$. In this case the corrected
$\Lambda(\vec x, \vec y)$ is
\be
\tilde \Lambda(\vec x, \vec y)=\sum_n \psi_n(\vec x) sign
\varepsilon_n f (|\varepsilon_n| T_g) \psi^+_n(\vec y)
\ee
where from (A.4.7) the function $f$ is (the exact form is given in
(A.4.5)
\be
f(|\varepsilon_n|T_g)=\left \{
\begin{array}{ll}
1-\frac{2|\varepsilon_n|T_g}{\sqrt{\pi}},&|\varepsilon_n|T_g\ll 1\\
\frac{1}{\sqrt{\pi}|\varepsilon_n|T_g},
&|\varepsilon_n|T_g\gg 1
\end{array}
\right.
\ee
On can check that the presence of this function in the sum (92) and
in the integral (98) does not change the main result: the appearance
of the smeared--off $\delta$--function with the same range parameter
(99), but instead of the function $\frac{2}{\pi} K_0(a)$ one obtains
another smeared $\delta$--function $K(a)$, finite at $a=0$
and exponentially decaying at large $a$. Therefore again one obtains
local at large $r$ interaction -- linear confinement for light
quarks.

More important is the change in the sum (119) which now should be
replaced by
\be
<\psi\bar \psi>_M=-\frac{N_c}{2\pi}\sum^{\infty}_{n=0}[(A^-_n)^2-
(B^+_n)^2]f(|\varepsilon_n|T_g)
\ee
Using the asymptotics (142) and (128), one can see that the sum is
now diverging only logarithmically, and this weak divergence may be
cured by the nonlocality caused by the rest $p_4$ dependence of
$M(p_4,\vec z, \vec w)$.

As a result one obtains an estimate
\be
<\psi\bar \psi>\cong - N_c\frac{\sigma\lambda}{T_g} ln
\frac{\varepsilon_{max}}{\varepsilon_0}
\ee
where $\lambda$ is the numerical factor, $\varepsilon_{max}\sim
1/T_g$. For $T_g\sim 0.2$fm  and numerical factors from (128) one
obtains a correct order of magnitude for the chiral condensate. A
more detailed calculation of $<\bar \psi \psi>$ in the quasiclassical
method is now in progress and will be published elsewhere.

Note, that the $N_c$ dependence in (119) and (143) is reproduced
correctly -- the chiral condensate should be proportional to $N_c$ in
the limit $N_c\to \infty$.

Since $\sigma \sim D(0) T_g^2\sim <F^2(0)>T_g^2,
$ one obtains for the chiral condensate parametrically
(up to a numerical constant)
\be
<\psi\bar \psi>=-\frac{\alpha_s}{\pi}<F^2(0)>T_g
\ee
Thus the chiral condensate diverges  in the "string limit of QCD",
i.e. when $\sigma=const,$ and $T_g\to 0$. In this limit the width of
the QCD  string tends to zero [21]. In the realistic case, i.e. for
finite $T_g$, and finite gluonic condensate, the $<\psi\bar \psi>$ is
also finite, and inserting in (145) the standard value [22]
$\frac{\alpha_s}{\pi}<F^2(0)>=0.012 GeV^4$, and $T_g=1 GeV^{-1}$ [23],
one obtains $<\psi\bar \psi>\approx - (250 MeV)^3 const$, with const
of  the order of one.

To understand better the source of divergence of the chiral
condensate at $T_g\to 0$ and to prove  the finiteness of $<\bar \psi
\psi>$ for the exact equations (15,16), let us look for the solution
of (16) in the form
\be
 S(x,y)=-i\hat \partial_x f(x,y)+g(x,y) \ee
Insertion of (146) into (16) yields for $m=0$.
\be
-\partial^2_x f(x,y)-i\hat \partial_x g(x,y)-i\int  M(x,z) (-i\hat
\partial_z f(z,y)+g(z,y)) d^4z=\delta^{(4)}(x-y)
\ee
where
$$
iM(x,z)=J^E(x,z)(-i\partial_4\gamma_4 f+g+i\vec \partial \vec \gamma
f)+
$$
$$
+J^B(x,z)[+3i \partial_4 f\gamma_4+3g+ i\vec \partial
\vec \gamma f-2i n_z\vec
\partial f\hat n_x-(g+i\hat\partial f)\hat n_z\hat n_x]=
$$
\be
= iM^{(1)}_{ik}\partial_i\gamma_k f(x,z)+M^{(2)}(x,z)g(x,z)
\ee
To satisfy (147), $f(x,y) $ should be singular and one can represent
$f$ as
\be
f(x,y) =\frac{1}{4\pi^2(x-y)^2}+\tilde f(x,y)
\ee
   and
   \be
\partial^2_x \frac{1}{4\pi^2(x-y)^2}=\delta^{(4)}(x-y)
\ee
As a result one obtains two equations
$$
-\partial_x^2\tilde f(x,y)=
\frac{1}{4} tr\int[M^{(1)}_{ik}\gamma_k\partial_i
f(x,z)\gamma_{k'}\partial_{k'} f(z,y)+
$$
\be
+M^{(2)}(x,z) g(x,z) g(z,y)] d^4z
\ee
\be
-i\partial_{\mu}g(x,y)=i\frac{1}{4} tr \gamma_{\mu}
\int [ M^{(1)}_{i\mu}\partial_i f(x,z) g(z,y)
- M^{(2)} (x,z) g(x,z)
\gamma_{\nu}\partial_{\nu}f(z,y)]d^4z
\ee
One can identify on the r.h.s. of (151) the most singular term, which
yields for $\tilde f$ at small $x,y$
\be
\tilde f (x,y) \sim ln |x-y| (x^2+xy)
\ee
where we have the property of $J^E, J^B$ at small $x,y$
\be
J^E, J^B\sim (x_{\mu}y_{\mu}) const
\ee
From (152) one can conclude that $g(x,y)$ is nonsingular at small
$x,y$:
\be
g(x,y)=g_0+g_1(x^2+y^2)+g_2 xy +...
\ee
Hence
\be
<\bar \psi \psi> =4N_c g(0,0)=4 N_c g_0
\ee
is nonsingular and finite for the solution of the full equations
(147).

In the small $T_g$ limit,  considered in the last two sections, when
the kernel $M$ is becoming local and yields the linear scalar
potential at large distances the situation changes. One can do the
same analysis as done above, Eqs. (146-152), and obtain for the local
time--independent kernel $M=\sigma|\vec x|\delta^{(3)}(\vec z-
\vec y))$
\be
\lim_{|\vec x|\to 0} g(x,0) =-\frac{i\sigma}{|\vec x|},
\ee
and when the time-nonlocality term is used in $M$, $M\sim exp
(-\frac{(h_4-h'_4)^2}{4T_g^2})$, $g(x,0)$ is less singular
$g(x,0)\sim ln |\vec x |, x\to 0$ but still diverges logarithmically.
These properties explain the divergence of the sum (119) in the
approximation (128), corresponding to (157), and the logarithmic
divergence of the sum (143) corresponding to the time--nonlocal case
(with the account of the $p_4$--dependence). Our conclusion is that
the accurate computation of the chiral condensate requires exact
solution of the equations (15), (16).

\section{Chiral symmetry breaking: zero modes $vs$ field correlators}

We have seen in previous sections that the nonlinear equations (15),
(16) give rize to the phenomenon of CSB, which reveals itself in our
problem in two ways: i) it provides scalar confining interaction for
the light quark ii) there appears a standard chiral condensate
$<\bar \psi \psi>$.

A natural question arises at this point: a folklore understanding of
CSB is that it is due to quasizero global quark modes in the vacuum.
An exact relation [24] exists, which connects chiral condensate to the
density $\nu(\lambda)$ of quasizero modes in the vacuum at
$\lambda\la m$, and in the chiral limit $(m\to 0)$ one has [24]
 \be
 <\bar \psi \psi>=-\frac{\pi\nu(0)}{V_4}
 \label{7.1}
 \ee
 Here $\lambda$ is an eigenvalue of the 4d Euclidean equation for the
 quark in the vacuum field $A_{\mu}$
 \be
 -i\hat D   \psi_n(x) = \lambda_n\psi_n(x)
 \label{7.2}
 \ee
 The density $\nu(\lambda)d\lambda$ is the averaged over all fields
 $\{A_{\mu}\}$ number of the states $\lambda_n$ per interval
 $d\lambda$.

 It is a popular belief that the quasizero modes necessary for CSB
 due to (\ref{7.1}) are descendant from the local zero modes on the
 topological charges (instantons or dyons), and their density is
 therefore proportional to the density of instantons (dyons). There
 are the instanton model [7,8] and  the dyon model [9] of the QCD
 vacuum, which explain CSB in this way.

 Whether these models are realistic or not, is the open question, but
 the Banks-Casher relation (\ref{7.1}) holds independently of that,
 and if the method of the present paper proves CSB due to the field
 correlators (even in the Gaussian approximation), one should explain
 the origin of the quasizero modes in (\ref{7.1}).

 To do this we consider first the case of Abelian fields. As was
 stressed above in the paper, CSB is due to the correlator $D(x)$,
 and the latter in the Abelian case can be connected to the
 correlator of magnetic monopole currents [1,2]
 \be
 <\tilde j_{\beta}(x) \tilde j_{\delta}(y)> =
 (\frac{\partial}{\partial x_{\alpha}}
 \frac{\partial}{\partial y_{\alpha}}
 \delta_{\beta\delta}-
 \frac{\partial}{\\partial x_{\beta}}
 \frac{\partial}{\partial y_{\beta}} ) D(x-y)
 \label{7.3}
 \ee

 In the nonabelian case one can use the Abelian projection method
 (APM) [25], to separate in the field $A_{\mu}$ and field strength
 $F_{\mu\nu}$ the monopole and photon part, and the part of "charged
 gluons". The latter contributes around 10\% to the effective action.
  In this case one can connect the monopole current obtained by
 APM with the correlator $D$ as in (\ref{7.3}).

 Now for each magnetic monopole (or dyon) there is an infinite number
 of fermion zero modes, proportional to
 the length $T$ of the world
  line of the monopole. Therefore the density of
 fermion zero modes per unit 3d volume and per unit of length along
 the world line is exactly $\frac{\nu(0)}{V_4}$, as in  (\ref{7.1}).
 On the other hand one can estimate this density from the 3d density
 of magnetic monopoles, which can be obtained from $<\tilde
 j_{\beta}(\vec x, x_4)\tilde j_{\beta}(0,x_4)>$ integrating over
 $d^3\vec x$. (The correlator $<\tilde j(x)\tilde j(0)>$ estimates
 probability of finding a monopole at the point $x$, if there is one
 at $x=0$. Integrating over $d^3\vec x$ one finds the probability of
 having a monopole at $x=0$, while another is anywhere fixing $x_4$
 means that the probability refers to a given moment. We assume that
 one magnetic monopole yields one quasizero fermion mode per unit
 length of its world line -- it is true for an isolated monopole, and
 we extrapolate this relation to the QCD vacuum as a whole.) Hence
 one gets an order of magnitude relation
 \be
 \frac{\nu(0)}{V_4}\sim \int d^3x<\tilde j(\vec x, x_4)\tilde
 j(0,x_4)>\sim D(0) T_g
\label{7.4}
\ee
 where we have assumed for $D(x)$ the form
 $D(x)=D(0)f(\frac{x}{T_g})$ and $f(y)$ is an exponential or Gaussian
 with $f(0)=1$.

 Finally, taking into account that $D(0)\sim g^2<F(0)F(0)>$ one
 obtains
 \be
 <\bar \psi\psi>\approx -\frac{g^2}{4\pi}<F^2(0)>T_g
 \label{7.5}
 \ee
 This estimate coincides with our result obtained from the
 quasiclassical calculation in the previous section. Numerically
 (\ref{7.5}) is -- $(300 MeV)^3$, i.e. a reasonable order of
 magnitude. Thus the very existence of the "wrong" correlator $D(x)$,
 violating Abelian Bianchi identity may bring about monopole currents
 and ascociated with those zero modes.

\section{The contribution of higher--order correlators}

The term in the cluster expansion of the effective action,
proportional to the connected average of $\ll A_{\mu_1}(x^{(1)})...
A_{\mu_n}(x^{(n)})\gg$ contributes to the operator $M$ the quantity
$$
iM^{(n)}(x^{(1)},...x^{(n)})=\gamma_{\mu_1}S(x^{(1)},x^{(2)})
\gamma_{\mu_2}...\gamma_{\mu_{n-1}}
S(x^{(n-1)},x^{(n)})\gamma_{\mu_n}\times
$$
\be
\times N^{(n)}_{\mu_1... \mu_n}(x^{(1)},...x^{(n)})
\label{8.1}
\ee
where we have defined
$$
 N^{(n)}_{\mu_1... \mu_n}= \int^{x_1}_0 d\xi^{(1)}_{\nu_1}
  \int^{x_2}_0 d\xi^{(2)}_{\nu_2}...
  \int^{x_n}_0 d\xi^{(n)}_{\nu_n}
  \alpha
  (\xi_{\nu_1})...
  \alpha
  (\xi_{\nu_n})\times
  $$
  \be
  \ll F_{\nu_1\mu_1}(\xi^{(1)})...
   F_{\nu_n\mu_n}(\xi^{(n)})\gg
   \ee
   and $  \alpha(\xi_4)=1,          ~~
    \alpha(\xi_i^{(k)})=
    \frac{\xi_i^{(k)}}{x^{(k)}_i},~~
    i=1,2,3; k=1,...n.$

    One  can identify in cumulant $\ll...\gg$ the part similar to
    $D$, i.e. violating the Abelian Bianchi identity, namely
    for even  $n$
  \be
  \ll F_{\nu_1\mu_1}(\xi^{(1)})...
   F_{\nu_n\mu_n}(\xi^{(n)})\gg =
   \prod_{i,k}(\delta_{\nu_i\nu_k}\delta_{\mu_i\mu_k}-
   \delta_{\nu_i\mu_k}\delta_{\nu_k\mu_i})D^{(n)}(\xi^{(1)},...
   \xi^{(n)})
   \ee

   Assuming for $D$ the Gaussian form (the result which follows does
   not depend on that assumption modulo numerical factors)
   \be
   D(u^{(1)},...u^{(n)})= D_0 exp
   (-\frac{1}{4T^2_g}\sum^n_{k=2}(u^{(k)}-u^{(k-1)})^2)
   \ee
   one obtains at large $\vec x^{(i)},i=1,...n, |\vec x^{(i)}|\gg
   T_g, |\vec x^{(i)}-\vec x^{(k)}|\la T_g, i,k =1,...n$.
     \be
      \lim
      \begin{array}{c}
      {N^{(n)}}\\
      {|\vec x^{(i)}|\to \infty}
      \end{array}
      = const |\vec x|\prod
   (\delta_{\mu_i\mu_k...}) \ee

   Now one can proceed in exactly the same way as we did for the
   bilocal
   Gaussian correlator, namely one performs Fourier transformation in
   $x_4^{(n)}$ and for small $p_4^{(n)} (p_4^{(n)} T_g\ll 1)$ one
   obtains the integral (we omit $\gamma$-matrices and Kronecker
   symbols for simplicity)
   $$
   M^{(n)}(p_4^{(n)}\approx 0, \vec x^{(1)},... \vec x^{(n)})\sim
   N^{(n)}\int S(\tilde p_4^{(1)})...
    S(\tilde p_4^{(n-1)})\frac{d\tilde p_4^{(1)}}{2\pi}
    \frac{d\tilde p_4^{(n-1)}}{2\pi}=
    $$
    \be
    =
    N^{(n)}\prod^{n-1}_{k=1}\Lambda_1(\vec x^{(k)}, \vec x^{(k+1)})
    \ee
    where $\Lambda_1(\vec x, \vec y)$ is the same function as
    obtained in section 5, Eq.(81). Hence $\Lambda_1$ is a moderated
    $\delta$--function, which puts all $\vec x^{(i)}$ in the vicinity
    of each other, and $M^{(n)}$ is the scalar function linearly
    growing with $|\vec x|\sim |\vec x^{(n)}|$.  Thus all
    conclusions about CSB and confinement of light quarks, which have
    been made based on the Gaussian correlator $N^{(2)}$, hold true
    also for any even order correlator $N^{(2k)}$, provided it
    contains the $D$--type term $D^{(n)}$.

    Now the question arises: what is the contribution of all the sum
    $\sum^{\infty}_{n=1} M{(2k)}$, and what is the result for the
    exact QCD vacuum?

    Here are two possible patterns.

    i) In the first case the higher order correlators are suppressed
    as compared to the Gaussian one.
     This may be true for the  real QCD vacuum, and there are at
     least two evidences for the suppression of higher, $n\geq 4 $,
     correlators.
     One fact concerns the behaviour of the static potential for
     static quarks in higher SU(N) representations. The gaussian
     correlator yields the quadratic Casimir operator as the
     coefficient on the potential, while higher correlators induce
     additional group structure as well [2]. The numerous lattice
     calculation yield the firm evidence for the quadratic Casimir
     operator (see [2] for discussion and refs.), while other
     structures are not seen.

     Another fact is the study of the QCD string profile, i.e. the
     field distribution inside the string. This profile computed from
     the Gaussian correlator previously defined on the lattice [23],
     coincides with the one, obtained by the lattice Monte-Carlo data
     [21]. Again no sign of the contribution of higher correlators.

     Thus so far one can accept the hypothesis of the dominance of
     the Gaussian correlator, the corresponding picture is called the
     Gaussian stochastic vacuum [1,26].

     ii) In the second pattern all higher correlators are important.
     One typical example, when this  happens, is the dilute instanton
     gas model [7,8].
     The effective Lagrangian looks like very similar to our starting
     expression, with the only exception: instead of averaging the
     product $<A(1)...A(n)>$ over vacuum fields and expressing  it
     through the field correlators as in $N^{(n)}$, one averages the
     instanton field $A(i)$ over positions and color orientations of
     instantons. The result for the effective Lagrangian at large
     $N_c$ was obtained in [27]
     $$
     {\cal{L} } =\sum^{\infty}_{n=2}\frac{N}{2nV_4N_c^n}\prod^n_{k=1}
     \psi^+_{\alpha_k}(p_k)\gamma_{\mu_k}\psi_{\beta_k}(p_k-q_k)
     \frac{dp_kdq_k}{(2\pi)^8}
     \times
     $$
     \be
     (2\pi)^4\delta(\sum^n_{i=1}q_i)tr(\prod^n_{j=1}A^I_{\mu_j}(q_j))
     \prod \delta_{\alpha_i,\beta_{i-1}}+I\leftrightarrow \bar I
     \label{8.7}
     \ee
     Here $A^I$ is the standard instanton vector potential
     in the singular gauge, $N$ is
     the number of instantons in the volume $V_4$ and $I\to \bar I$
     implies a sum over antiinstantons.

     From (\ref{8.7}) one easily finds the contribution of all terms
     with the product of $n$ instanton fields (which is an
     equivalent to the $n$-th  field correlator), this has been done
     in [27] with the result
$$
M=i\sum^{\infty}_{n=2}\frac{N}{2V_4N_c}(\gamma_{\mu_1}S(p_1-q_1)
\gamma_{\mu_2} ... S(p_n-q_n)\gamma_{\mu_n}\times
$$
\be
\times
     (2\pi)^4\delta(\sum^n_{i=1}q_i)tr(\prod^n_{j=1}A^I_{\mu_j}(q_j))
    +I\to \bar I
    \label{8.8}
    \ee
      In the sum (\ref{8.8}) all terms are important, and the sum can
      be computed explicitly [27] resulting in the equation, which was
      found earlier by another method [28].
\be
M(p)=i\frac{N}{2N_cV_4}<p|(\bar
S-(A^I)^{-1})^{-1}|p>+I\leftrightarrow \bar I
\label{8.9}
\ee
Here $\bar S$ is the averaged propagator
\be
\bar S=\frac{1}{\hat p-iM(p)}
\label{8.10}
\ee
The nonlinear equation (\ref{8.9}) can be studied  in the low
density limit,\\
 $\frac{N\rho^4}{V_4N_c}\ll 1, N_c\to \infty$, where
$\rho$ is the average size of instantons. One can argue, that the
nonzero $M(p)$, i.e. the CSB phenomenon, occurs for any
density [27,28].

At the same time, as mentioned above, the confinement and hence the
string between the light quark and the heavy antiquark here is
missing, since instantons do not ensure confinement [2].

Thus the effect of higher correlators in the case of instanton gas
model is twofold: on one hand higher correlators cancel in the sum
for the string tension and hence destroy confinement, while it was
present in the Gaussian approximation [1].
Remarkably this cancellation occurs only for integer topological
charges as is the case for instantons, and is absent for magnetic
monopole (dyonic) charges [29].

On another hand the higher correlators sum up in the new nonlinear
equation (\ref{8.9}) which produces CSB as well as in the case of the
purely Gaussian correlator, Eqs.(15),(16).

\section{Discussion and prospectives}

We have studied in this paper the simplest gauge--invariant system,
containing a light quark bound together with a heavy quark. This
gauge--invariant setting allows for the appearance of the QCD string
and the main subject of the paper is the influence of the string on
the dynamics of one light quark at the end of this string. The vacuum
correlator method has enabled us to introduce and to describe the
string in the model--independent way, and the most part of the paper
is denoted to the discussion of the bilocal Gaussian correlator. But
this is because that simplest correlator already ensures the
appearance of  the string, and the most general correlators,
discussed in section 8, do the same job, and as it was shown there,
bring about the CSB in the same way as the Gaussian correlator.

There have been used several simplifications. The first one is the
large $N_c$ limit, which allows to neglect additional quark loops
(quenching approximation) and to  factorize averages of a product of
loop integrals (e.g.in Appendix 1). Correspondingly the problem
effectively reduces to the one-flavour problem, since the diagrams
taken into account are the one--quark flow diagrams, (however with
backward motion in time). This allows us to calculate the chiral
condensate, which essentially  constitutes a closed light quark loop,
and by choosing in $S(x_4-y_4,\vec x,\vec y)$ both times $x_4,y_4$
and coordinates $\vec x, \vec y$ coincident, one disconnects the
light quark loop from the heavy antiquark.

So far the problem of several light quark flavours was not considered
in the paper. As the next step of the same formalism one can
introduce the $q\bar q$ Green's function and write
for it  the nonlinear
equation similar to equations of the present paper, but taking
into account both light quarks.
Evidently the same formalism works
for the baryon -- $3q $ -- state. This will be a subject of a future
work.

The main idea of the paper and of this future formalism is the
selfinteraction of quark, which brings about nonlinear interaction
kernel in the Dirac--type equation, and this nonlinearity breaks
chiral symmetry. In this way the spontaneous symmetry breaking
happens, and this seems to be a much more general phenomenon, than
was assumed before. In particular, the same is the pattern of CSB in
the NJL model. It is likely that the very concept of nonperturbative
phenomena occurs in the same way -- as the spontaneous breaking of
scale invariance, and that happens via new solutions of nonlinear
equations for gauge--invariant correlators. This is a fascinating way
for the construction of the nonperturbative theory of gauge fields,
in particular QCD [30].

Coming back to the topic of the present paper, the main task left
uncompleted is the exact calculation of the chiral condensate. The
realistic estimates based on WKB have been given in section 6, but
the logarithmically divergent sum has been cut off in a reasonable,
but approximate way. To calculate chiral condensate exactly, one
should solve the nonlinear equations (15),(16) without local
approximation, used in section 6  and valid actually at large $|\vec
x|$. This will be the subject of future publications.

The author is grateful to H.Leutwyler for stimulating discussions in
June of 1994 and kind hospitality at Bern ITP. His remarks that CSB
may be caused by the Gaussian correlators have
encouraged the present
work.

The large part of this work has been done when the author was
visiting ITP of Utrecht University in January of 1997 supported by
the  INTAS grant  93-79.

The author is grateful to G.'tHooft and J.A.Tjon for valuable
discussions, and to J.A.Tjon for cordial hospitality. The author
thanks ITEP seminar members and particular A.B.Kaidalov, V.A.Novikov,
Yu.S.Kalashnikova, V.S.Popov and K.A.Ter--Martirosyan for useful
discussions. A partial   support of RFFR grants 96-02-19184 and
95-048-08 is gratefully acknowledged.
\newpage

\underline{\bf Appendix 1}
\setcounter{equation}{0} \def\theequation{A1.\arabic{equation}}

 \vspace{1cm}

{\large {\bf Derivation of the effective Lagrangian (12)}}

\vspace{1cm}

In this Appendix two ways of derivation of (12) are described:

1) a simple way using the gauge (5), (6)

2) a gauge--invariant derivation. In both cases we assume that $N_c$
is large.

We start with the first one. Writing $A_{\mu}^{ab}$ with fundamental
color indices as
$$
A^{ab}_{\mu}=t^A_{\mu}, A=1,...,N^2_c-1,a,b=1,...N_c
  $$
  \be
  tr t^At^B=\frac{1}{2}\delta_{AB}
  \ee
  one has in the gauge (5),(6)
  \be
  A_{\mu}^A=\int \alpha(u)du F_{\nu\mu}^A(u)
  \ee
  Hence for the averaged product one obtains
  \be
  <A^{ab}_{\mu}(z)A^{cd}_{\mu'}(w)>=
  t^A_{ab}t^B_{cd}
  \int \alpha(u)du_{\nu}
  \int \alpha(u')du'_{\nu}
  < F_{\nu\mu}^A(u)
   F_{\nu'\mu'}^B(u')>
   \ee
   Using now color neutrality of the vacuum one has
   \be
g^2<F^A(u)F^B(u')>=\frac{\delta_{AB}}{N_C^2-1}
g^2<F^C(u) F^C(u')>
\ee
The r.h.s. of the Eq. (A~1.4) can be expressed through $D$, defined
in (10), namely
               \be
g^2<F^C(u) F^C(u')>=2N_cD(u-u'),
\ee
       since $<trF(u)F(u')>=\frac{1}{2}
<F^C(u) F^C(u')>$. Finally using the relation
\be
t^A_{ab}t^A_{cd}=\frac{1}{2}\delta_{ad}\delta_{bc}-\frac{1}{2N_c}\delta_{ab}
\delta_{cd}
\begin{array}{l}
{\longrightarrow}\\
{N_c\to\infty}
\end{array}
\frac{1}{2}\delta_{ad}\delta_{bc}
\ee
One obtains
\be
g^2<A^{ab}_{\mu}(z)A^{cd}_{\mu}(w)>
=\frac{N_c}{N_c^2-1}
\delta_{ad}\delta_{bc}
  \int \alpha(u)du
  \int \alpha(u')du'
  D(u-u')
  \ee
  At large $N_c$ Eq.(A~1.7) being inserted in Eq(4), reproduces (12).

  2) In gauge-invariant version of relations (5), (6) one can insert
  there parallel transporters $\Phi(z,u)$ as follows
  \be
  A^{ab}_{\mu}(\vec z, z_4) =
  \int \alpha^{(\mu)}(u) du_i \Phi^{aa'}(\vec z,z_4; \vec u, z_4)
  F_{i\mu}^{a'b'}(\vec u, z_4)
   \Phi^{b'b}(\vec u,z_4; \vec z, z_4)
  \ee
  the same representation one can use for $A_{\mu'}(w)$.

  Now one can insert in both expressions (A~1.18) for $A_{\mu}(\vec
    z, z_4)$ and $A_{\mu'}(\vec w, w_4)$ the unity on both sides of
  $F$, namely
  \be
   \Phi(\vec u,z_4; 0, z_4)
   \Phi(0, z_4; \vec u,z_4) =1,~~ {\rm {for}}~~ A_{\mu}(\vec z, z_4)
   \ee
   and
   \be
   1=
   \Phi(\vec w,w_4; 0w_4)\Phi(0,w_4; 0z_4)
   \Phi(0,z_4;0w_4) \Phi(0,w_4; \vec w, w_4)
   \ee
    As a result one can write
  \be
  A^{ab}_{\mu}(\vec z, z_4) =
  \int \alpha^{(\mu)}(u) du_i \Phi^{aa'}(\vec z,z_4;0, z_4)
  F_{i\mu}^{a'b'}(\vec u, z_4;0,z_4)
   \Phi^{b'b}(0,z_4; \vec z, z_4)
  \ee
  \be
  A^{cd}_{\mu'}(\vec w, w_4) =
  \int \alpha^{(\mu)}(u') du'_{i'} \Phi^{cc'}(\vec w,w_4;0, z_4)
  F_{i'\mu'}^{c'd'}(\vec w, w_4;0,z_4)
   \Phi^{d'd}(0,z_4; \vec w, w_4)
  \ee

        Now both $F^{a'b'}$ and $F^{c'd'}$ are referred to one point
        $(\vec 0,z_4)$, namely we have defined
$$
F^{a'b'}(\vec u,z_4;0z_4)=
(\Phi(0,z_4;\vec u,z_4) F(\vec u,z_4) \Phi(\vec u,z_4;0,z_4))^{a'b'}
$$
  \be
F^{c'd'}(\vec w,w_4;0z_4)=
(\Phi(0,z_4;\vec w,w_4) F(\vec w,w_4) \Phi(\vec
w,w_4;0,z_4))^{c'd'}
 \ee
  The average value of $<A_{\mu}A_{\mu'}>$ (cf
  Eq. (7) of the main text) is expressed through the product of $F$
  in Eq. (A~1.13), and both $F$ there are gauge tranformed as
  \be
     F(\vec x,x_4; 0,z_4)\to \Omega^+(\vec 0,z_4)F(\vec x, x_4;
     0,z_4) \Omega(\vec 0, z_4)
     \ee
      and therefore the averaging of
  two $F$ with this property  produces the gauge invariant,
  namely $$ g^2<F^{a'b'}_{i\mu}(\vec u,z_4;0z_4)
  F^{c'd'}_{i'\mu'}(\vec w,w_4;0z_4)>=
$$
\be
=\frac{\delta^{a'd'}\delta^{b'c'}}{N_c^2} g^2<tr(F(\vec
u,z_4;0z_4)  F(\vec w,w_4;0z_4))>=
\ee
$$
=\frac{\delta^{a'd'}\delta^{b'c'}}{N_c}
(\delta_{ii'}\delta_{\mu\mu'}-\delta_{i\mu'}\delta_{i'\mu})D(u,w)
$$
Therefore one obtains
$$
<A^{ab}_{\mu}(\vec z, z_4) A^{cd}_{\mu'}(\vec w, w_4)>=
  \int \alpha^{(\mu')}(u') du'_{i'}
  \int \alpha^{(\mu)}(u) du_i\times
  $$
  \be
  \times
   \Phi^{aa'}(\vec z,z_4;0 z_4)
   \Phi^{a'd}(0,z_4; \vec w, w_4))
   \Phi^{cc'}(\vec w,w_4;0 z_4)
   \Phi^{c'b}(0,z_4; \vec z, z_4)
  \times
  \ee
  $$
  \times \frac{1}{N_c^2}<tr
  F_{i\mu}(\vec u, z_4;0z_4)
  F_{i'\mu'}(\vec w, w_4;0z_4)>
  $$
  The insertion of this expression into  Eq. (4) yields for the term
   bilinear in $A$
   $$
   {\cal{L}}_{eff}=-\frac{1}{2N_c}\int dz\int dw
   (
   \psi^+_a(z)\gamma_{\mu}\psi_b(z))(\psi^+_c(w)\gamma_{\mu'}\psi_d(w))\times
   $$
   \be
   (\delta_{\mu\mu'}\delta_{ii'}-\delta_{i\mu'}\delta_{i'\mu})
   J^{\mu\mu'}_{ii'}(z,w)
   \ee
   Here we have defined
   \be
   \Psi_{ad}=(\Phi(\vec z, z_4;0z_4)\Phi(0,z_4,\vec w,
   w_4))_{ad}
   \ee

   Finally we note, that $A_{\mu}$ enters $\Psi,\Psi^+$,
   in the expression (A~1.17) which is assumed to be
   averaged over $A_{\mu}$, and indeed, the averaged
   correlator (A~1.15) enters $J^{\mu\mu'}_{ii'}$, which
   contains $D(u-u')$. Therefore the $\Psi,\Psi^+$ in
   (A~1.17) are to be understood as averaged in the
   invariant combinations:
   \be
   <\psi^+_a(z)\Psi_{ad}(z,w)\psi_d(w)>_A,
   \ee
   \be
   <\psi_b(z)\Psi_{cb}^+(z,w)\psi_c(w)>_A,
   \ee
   Note also, that the invariant combinations (A~1.15),
   (A~1.19) and (A~1.20) can be averaged separately
   inside the common expression ${\cal{L}}_{eff}$ with
   accuracy of $1/N_c^2$ for large $N_c$.

   Finally, one can see that $D(u,w)$ in (A~1.15) is
   defined in a gauge--invariant way, but depends not
   only on $u,w$, but also on the contour connecting
   those points. However one can see in the main text,
   that $|z_4-w_4|\sim T_g$ and in the limit $T_g\to 0$
   one can replace the complicated $\prod$-shaped contour
   in (A~1.15) by a straight line connecting $u$ and $w$,
   with an accuracy of $T_g^2$. For the straight--line
   contour the average in (A~1.15) and hence $D(u,w)$,
   reduces to the function $D(u-w)$. This is what is
   written in Eq.(10) of the main text, and used
   throughout the paper.

\newpage

\underline{\bf Appendix 2}
\setcounter{equation}{0} \def\theequation{A2.\arabic{equation}}

 \vspace{1cm}

{\large {\bf The term $\Delta^{(1)}$ and its role in the
deconfinement transition.}}

\vspace{1cm}

 The term $\Delta^{(1)}$  defined in (10) has the representation
 $$
 \Delta^{(1)}=
 \frac{1}{2}\{\frac{\partial}{\partial
 u_i}[(u-u')_{i'}\delta_{\mu\mu'}-( u-u')_{\mu'}\delta_{\mu i'}]+
  $$
  \be
 +\frac{\partial}{\partial u_\mu}[(u-u')_{\mu'}\delta_{ii'}-(
 u-u')_{i'}\delta_{i\mu '}]\}D_1(u-u'),
 \ee
 where derivatives act also on $D_1$, so that the whole form (A~2.1)
 is a total derivative, and hence in the double integral $d^2ud^2u'$
 entering in the calculation of the Wilson loop [1] it contributes
 not to the area law term \\$<W>\approx exp(-\sigma$ area), but rather
 to the perimeter term $exp CL$ where $L$ is the length of the
 contour.

 The contribution of $\Delta^{(1)}$ to the effective Lagrangian (12)
 is obtained replacing $(\delta\delta-\delta\delta) D$ in (13) by
 $\Delta^{(1)}$, given in (A~2.1). As a result one obtains $\tilde
 J^{\mu\mu'}_{ii'}$ consisting of a sum of 4 terms, which correspond
 to 4 successive terms in $\Delta^{(1)}$, Eq. (A~2.1)
$$
\tilde J_{\mu\mu'}(z,w)\equiv \int^z_0 du_i \alpha_{\mu}(u)
\int^w_0du'_{i'} \alpha_{\mu'}(u')\Delta^{(1)}_{ii',\mu\mu'}(u-u')=
$$
\be
=^1\tilde J_{\mu\mu'}+^2\tilde J_{\mu\mu'}+^3\tilde
J_{\mu\mu'}+^4\tilde J_{\mu\mu'}.
\ee
We assume for $D_1$ the same Gaussian form as for $D$, namely
\be
D_1(u)=D_1(0) exp (-u^2/4T_{g1}^2)
\ee
\be
^1 \tilde J_{44}(z,w) = T_{g1}^2D_1(0)
(e^{-\frac{(z-w)^2}{4T_{g1}^2}}
e^{-\frac{z^2}{4T_{g1}^2}}
e^{-\frac{w^2}{4T_{g1}^2}} +1)
\ee
\be
^1\tilde J_{kk'}=\int^z_0\alpha(u) du_i\int^w_0 \alpha(u')
du'_{i'}\frac{\partial}{\partial u_i}((u-u')_{i'}
D_1(u-u'))\delta_{kk'} \ee
 $$ ^2\tilde J_{ik}=
w_i\{\frac{1}{2}\int^1_0t'dt'\int^1_0 dt (z_k t-w_{k}t')D_1(zt-wt')-
$$
\be
-\frac{1}{2}\int^1_0(z_k-w_kt')D_1(z-wt')t'dt'\}
\ee
\be
^3\tilde J_{44}=\frac{1}{2} \int^z_0 du_i\int^w_0 du'_i
\frac{\partial}{\partial u_4}[(u-u')_4 D_1(u-u')]
\ee
\be
^3\tilde J_{kk!}=\frac{1}{2} \int^z_0\alpha(u) du_i\int^w_0\alpha
(u') du'_i \frac{\partial}{\partial u_k}[(u-u')_{k'} D_1(u-u')]
 \ee
\be
^4\tilde J_{\mu\mu'}=^2J_{\mu'\mu}(\vec z\leftrightarrow \vec w)
\ee
We start now to discuss the contribution of different   terms
(A~2.4-A~2.9) to the kernel $J(\bar q,\bar q') $ as in (20), and to
the dimensionless kernels $J^{\mu\mu'}_{ik}$ as in (26-27). If one
writes those in the form
\be
^s\tilde J_{\mu\mu}=\frac{(D_1(0) T^3_g)}{\alpha^3}
(\alpha T_g^2)^{\nu} f^s_{\mu\mu'}(Q,Q'),~~s=1,...,4
\ee
then the power $\nu$ for each term $^s\tilde J$ would characterize
its  importance in the limit $\alpha\to 0$, e.g. terms with $\nu=0$
might contribute to the final  equation (32), while the terms  with
$\nu>0$ are suppressed and vanish in the limit $\alpha\to 0$.

We start with $^1\tilde J_{44}$. The first three terms on the r.h.s.
of (A~2.4) yield $\nu=\frac{3}{2}$ and therefore are not important in
the limit $\alpha \to 0$ (i.e. at large distances). The last term on
the r.h.s. of  (A~2.4) is constant, and  a simple calculation yields
for that $\nu=0$, therefore one should look at the effect of this
term more closely. We shall show now that the constant term in
$\tilde J$ cannot yield the selfconsistent nonzero solution of
equations (20),(21) or (31), (32). To this end we insert the constant
$J(z,w)$ or equivalently $J(\bar q, \bar q')=const  \delta (\vec q)
\delta (\vec q')$ on the r.h.s. of (20), and realize that solutions
for $M,S$ should have the form
$$
M(p_4,\vec p, \vec p')=\mu (p_4,\vec p) \delta(\vec p+\vec p')
(2\pi)^3
$$
\be
S(p_4,\vec p, \vec p')=
\sigma (p_4,\vec p) \delta (\vec p+\vec p') (2\pi)^3
\ee
From (20),(21) one then gets an equation
\be
\mu( p_4,\vec p)= const \frac{\mu(p_4,\vec p)}{\sqrt{\vec
p^2+\mu^2(p_4,\vec p)}}
\ee
which has only the trivial solution $\mu\equiv 0$.

For (A~2.5) one gets $\nu(^2\tilde J_{ik})=1$, this term is
unimportant. The term $^3\tilde J_{44}$, Eq.(A~2.7) contains the full
time derivative, which being Fourier transformed as in (32) yields
additional factor $(P_4-P'_4)^2\alpha T_g^2$ which can be neglected.
For the rest kernels,  Eqs. (A~2.8) and (A~2.9), the corresponding
$\nu=1$.

Thus the analysis of the nonconfining correlator $D_1$, brings one to
the conclusion that it cannot yield a selfconsistent solution for the
scalar kernel $M$, and hence cannot break  chiral symmetry.

\newpage

\underline{\bf Appendix 3}
\setcounter{equation}{0} \def\theequation{A3.\arabic{equation}}

 \vspace{1cm}

{\large {\bf Properties of the kernel   $J_{ik}$
}}

\vspace{1cm}

As defined in (13), $J_{ik}$ is
$$
J_{ik}^{\mu\mu'}= \int^z_0 du_i \alpha_{\mu}(u)\int^w_0
\alpha_{\mu'}(u') du'_k D(u-u')=
$$
\be
= z_iw_k\int^1_0 dt \alpha_{\mu}(t)\int^1_0  dt' \alpha_{\mu'}(t')
D(\vec zt- \vec w t', z_4-w_4)
\ee
For the Gaussian form of $D$, one has
\be
D(\vec z t-\vec w t', z_4-w_4)= D(0) exp
(-\frac{(z_4-w_4)^2}{4T_g^2})  exp (-\frac{(\vec z  t-\vec w
t')^2}{4T_g^2})
\ee
The last factor can be rewritten as
\be
\eta(\vec z, \vec w) \equiv  (-\frac{1}{4T_g^2}[\frac{(\vec z+\vec
w)^2}{4}(t-t')^2+ \frac{(\vec z-\vec w)^2}{4} (t+t')^2+
\frac{(\vec z^2-\vec w^2)}{2}(t^2-t^{'2})])
\ee

The integral $\int dt dt'\eta (\vec z, \vec w)$ has different
asymptotics in 3 different regions of the $\vec z, \vec w$ space.

1)$
|\vec z-\vec w| \ll |\vec z+\vec w|, |\vec z-\vec w|\la T_g$

2)$
|\vec z-\vec w| \ll |\vec z+\vec w|, |\vec z-\vec w|\gg T_g$

3)$
|\vec z-\vec w| \sim |\vec z+\vec w|\gg T_g$

In the region 1) one has
\be
J^{44}_{ik}=z_iw_k D(0) e^{-\frac{(z_4-w_4)^2}{4T_g^2}}
\frac{4T_g\sqrt{\pi}}{|\vec z+\vec w|}(1+0(\frac{T_g}{|\vec z+\vec
w|}))
\ee
\be
J^{ik}_{ik}=\frac{1}{3} J^{44}_{ik}
\ee
In the region 2),3) it is convenient to rewrite the integral
$J^{44}_{ik}$ as
\be
J^{44}_{ik}=z_iw_k D(0) e^{-\frac{(z_4-w_4)^2}{4T_g^2}}
\frac{4T_g}{|\vec z+\vec w|}\frac{2T_g}{|\vec
z-\vec w|}\int ^{\xi_0}_0 d\xi
\int^{c\xi}_{-c\xi} d\lambda
e^{-\lambda^2-\xi^2-2\lambda\xi cos \chi}
\ee
where we have defined
\be
cos \chi= \frac{(\vec z^2-\vec w^2)}{|\vec
z-\vec w||\vec z+\vec w|},~~
c=\frac{|\vec z+\vec w|}{|\vec z-\vec w|},~~ \xi_0
= \frac{|\vec z-\vec w|}{2T_g}
\ee

 For $c\gg 1$ and $ \xi_0\ga 1$ one can expand the
 integration region in $\lambda$ to $(-\infty,
 \infty)$ and shift the variable as $\lambda+\xi
 cos \chi =\lambda'$, which yields for the last
 two integrals in (A~3.6)
 \be
 \sqrt{\pi}\int^{\xi_0}_0 d\xi e^{-\xi^2 sin^2\chi}
 =\left\{
 \begin{array}{ll}
 \frac{\pi}{2|sin \chi|},&\chi\neq 0, \xi_0|sin \chi|\gg 1\\
 \sqrt{\pi} \xi_0,&\xi_0|sin \chi|\la 1
 \end{array}
 \right.
 \ee

 Insertion of (A~3.8) into (A~3.6) finally yields in region (2)
 \be
 J^{44}_{ik}= z_i w_k D(0) e^{-\frac{(z_4-w_4)^2}{4T_g^2}}
 \frac{4T_g^2\pi}{\sqrt{|\vec z-\vec w|^2|\vec z+\vec w|^2-(\vec
 z^2-\vec w^2)^2}}
 \ee
 For $\xi_0|sin \chi |\la 1$  one using  (A~2.8) recovers the result
 (A~3.4).

 In the region 3) the integrals in (A~3.6) are of the order $0(1)$
 unless $ cos\chi =\pm 1$, and the estimate of $J^{44}_{ik}$ is
 typically therefore
 \be
 J^{44}_{ik}= z_i w_k D(0) e^{-\frac{(z_4-w_4)^2}{4T_g^2}}
 \frac{4T_g^2\pi}{|\vec z+\vec w||\vec z-\vec w|}
 f(|(\vec
 z+\vec w|/|\vec z-\vec w|)
 \ee
  where we have defined
  \be
  f=\frac{2}{\pi}\int^{\xi_0}_0 d\xi \int^{c\xi}_{-c\xi} d\lambda
  e^{-\lambda^2-\xi^2-2\lambda \xi cos \chi}
  \ee

  For $cos \chi =\pm 1$, i.e. when $\vec z $ and $\vec w$ are
  (anti)parallel and both large in modulus, one can choose coordinate
  $(\tau,\tau')$ in (A~3.2) such that
  $\tau=t-\frac{|\vec w|}{|\vec z|}t',\tau'$ orthogonal to $\tau$,
  and integration over $d\tau$ yields the factor $\frac{1}{|\vec
  z|}$, as in the case 1).

  Finally one can write for $J^{ik}_{ik}$, noting that
  \be
  tt'=4T_g^2(\frac{\xi^2}{|\vec z-\vec w|^2}-\frac{\lambda^2}{|\vec
  z+\vec w|^2})
  \ee
  therefore since in the integral (A~3.6) both $\lambda$ and $\xi$
  are of the order of one, hence one obtains in the regions 2),3),
  that
  \be
  J^{ik}_{ik}\sim \frac{T_g^2}{(\vec z\pm \vec w)^2} J^{44}_{ik}
  \ee

\newpage

\underline{\bf Appendix 4}
\setcounter{equation}{0} \def\theequation{A~4.\arabic{equation}}

 \vspace{1cm}

{\large {\bf Corrections to the  kernel (48) at finite   $T_g$
}}

\vspace{1cm}

  In the calculation of $M(p_4,\vec z,\vec w)$, Eq. (44), one should
             do the integral
\be
N(p_4)\equiv \int^{\infty}_{-\infty}\frac{ dp'_4 e^{-(p'_4-p_4)^2
             T_g^2}}{2\pi(p'_4-i\varepsilon_n)}
             \ee

 In section 4 the limit $T_g\to 0$ was used in which case the answer
 was given in (47). The limit however depends on the boundary
 conditions at infinity, and therefore should be done with care.
 Moreover we calculate in this appendix corrections due to finite
 values of $T_g$. To proceed one can represent $N(p_4)$ as follows
 \be
 N(p_4)=\int^{\infty}_{-\infty}\frac{dh_4e^{-\frac{h^2_4}{4T_g^2}
 -ih_4p_4}}{2\sqrt{\pi}T_g}\nu(h_4)
 \ee
 where
 \be
 \nu(h_4)\equiv \int \frac{
dp'_4 e^{p'_4h_4}}
{2\pi(p'_4-i\varepsilon_n)}=
\{
\begin{array}{ll}
i\theta(h_4)\theta(\varepsilon_n) e^{-|\varepsilon_n h_4|}, &
\varepsilon_n >0\\
-i\theta(-h_4)\theta(-\varepsilon_n) e^{-|\varepsilon_n h_4|}, &
\varepsilon_n <0
\end{array}
\ee
Inserting (A~4.3) into (A~4.2) one gets
\be
N(p_4)=\frac{i~sign(\varepsilon_n)}{2\sqrt{\pi}T_g}\int^{\infty}_0
dh_4
e^{-\frac{h^2_4}{4T_g^2}-h_4
\frac{\varepsilon_n}{|\varepsilon_n|}(\varepsilon_n+ip_4)}
\ee
Taking into account, that
\be
\int^{\infty}_0 dh e^{-\frac{h^2}{4T_g^2}-h\gamma}\left (
\begin{array}{l}
cos(p_4h)\\
sin(p_4h)
\end{array}
\right )
=
\ee
$$\left (
\begin{array}{l}
1\\
-i
\end{array}
\right )
\frac{\sqrt{\pi 4T_g^2}}{4}
\left\{
\begin{array}{l}
exp((\gamma-ip_4)^2T_g^2)[1-\Phi((\gamma-ip_4)T_g)]\\
\pm exp((\gamma+ip_4)^2T_g^2)[1-\Phi((\gamma+ip_4)T_g)]
\end{array}
\right.
$$
one obtains the following expansion of $N$ at small $T_g$,
\be
N(p_4)=\frac{i~sign(\varepsilon_n)}{2}(1-\frac{2|\varepsilon_n|T_g}
{\sqrt{\pi}}
+(\varepsilon_n^2-p_4^2)T_g^2-\frac{2p_4}{\sqrt{\pi}}(sign\varepsilon_n)
 T_g+0(T_g^3) \ee

In the opposite limit, when $|p_4|T_g\gg1$ or/and
$|\varepsilon_n|T_g\gg1$ one has from (A~
4.5)
\be
N(p_4)=
\frac{(i\varepsilon_n+p_4)}{2\sqrt{\pi}T_g(\varepsilon^2_n+p_4^2)}
\ee

\newpage

\underline{\bf Appendix 5}
\setcounter{equation}{0} \def\theequation{A5.\arabic{equation}}

 \vspace{1cm}

{\large {\bf The limit of the large mass $m$
 }}

\vspace{1cm}

In the equation for $S$
\be
(-i\hat \partial-im)S(h_4,\vec z,\vec w)-\int iM(h'_4,\vec z, \vec
 z')S(h_4-h'_4,\vec z', \vec w) dh'_4 d^3\vec z'=
 \delta(h_4)\delta^{(3)}(\vec z-\vec w)
 \ee
 for large $m\gg M$ one can neglect in the first approximation $M$
 and find the free Green's function $S_0$ for the massive quark,
 \be
 S_0(h_4,\vec z,\vec w)= (im-i\hat \partial) f(z,w)
 ,
 \ee
 \be
 f(z,w)=\int\frac{d^4pe^{ipx}}{(2\pi)^4(p^2+m^2)}=
 \frac{m}{4\pi^2 |x|}K_1(m|x|),\vec x=\vec z-\vec w
 \ee
 Since we shall treat differently $h_4=z_4-w_4$ and $\vec z- \vec
 w$, another representation is useful,
 $$
 S_0(h_4,\vec z, \vec w)=
 \frac{i}{2}\int \frac{d^3p}{(2\pi)^3}
 \frac{e^{i\vec p(\vec z-\vec w)}}{\sqrt{\vec p^2+m^2}}
 \{e^{-h_4\sqrt{\vec p^2+m^2}}\theta(h_4)(m+\gamma_4\sqrt{m^2+\vec
 p^2}-i\vec p\vec \gamma)
 $$
 \be
 +
 e^{h_4\sqrt{\vec p^2+m^2}}\theta(-h_4)(m-\gamma_4\sqrt{m^2+\vec
 p^2}-i\vec p\vec \gamma)\}
 \ee
 The form (A~5.4) allows the expansion in powers of $1/m$, which
 yields:
 \be
 S_0(h_4,\vec z, \vec w)= S_0^{(0)}(h_4,\vec z, \vec w) +\frac{1}{m}
 S_0^{(1)}(h_4,\vec z, \vec w)+...
 \ee
 where
 \be
 S_0^{(0)}(h_4,\vec z, \vec w)=\frac{i}{2}\delta^{(3)}(\vec z-\vec
 w)\{ e^{-h_4m}\theta(h_4)(1+\gamma_4)+
 e^{h_4m}\theta(-h_4)(1-\gamma_4)\}
 \ee

 Insertion of (A~5.6) into the expression for $M$
 \be
 iM(h_4,\vec z, \vec w)= J^{\mu}_{ik}(h_4,\vec z,\vec w)
 (\gamma_{\mu}S(h_4\vec z,\vec w) \gamma_{\mu} \delta_{ik} -
 \gamma_kS\gamma_i)
 \ee
 i.e. replacing  $S$ by $S_0^{(0)} $ yields
 \be
 iM(h_4,\vec z, \vec w)= \frac{i}{2}\tilde J(\vec
 z,\vec z) \delta^{(3)}(\vec z-\vec w) \{\theta
 (h_4) e^{-mh_4}[2-\frac{1}{3}(1-\gamma_4)]+
 \theta
 (-h_4) e^{mh_4}[2-\frac{1}{3}(1+\gamma_4)]\}
 \ee
 where we have defined
 \be
 \tilde J(\vec z,\vec z) =\vec z^2 \int^1_0\int^1_0 dt dt'D(\vec
 z(t-t'),h_4)
 \ee
 For the Gaussian form of $D$ one has at large $|\vec z|$:
 \be
 \tilde J(\vec z,\vec z) =\frac{\sigma|\vec
 z|}{T_g\sqrt{\pi}}e^{-\frac{h^2_4}{4T_g^2}}
 \ee

 Insertion of (A~5.8) into (A~5.1) yields
 $$
 (-i\partial_4\gamma_4-i\vec \gamma\vec \partial-im)S(h_4,\vec z,
 \vec w) -i\frac{\sigma|\vec z|}{2T_g\sqrt{\pi}}\int e^{-
 \frac{h^{'2}_4}{4T_g^2}} dh'_4\times
 $$
 $$
 \times
 \{\theta(h'_4)e^{-mh'_4}[2-\frac{1}{3}(1-\gamma_4)]+\theta(-
 h'_4)e^{mh'_4}[2-\frac{1}{3}(1+\gamma_4)]\}
 S(h_4-h'_4\vec z, \vec w)=
 $$
 \be
 =\delta(h_4)\delta^{(3)}(\vec z-\vec w)
 \ee
  To get rid of the $h_4$ dependence, we multiply both sides
  of (A~5.11) by $e^{mh_4}$ and integrate over $dh_4$.
  Defining
  \be
  \bar S(\vec z,\vec w)=\int^{\infty}_{-\infty} e^{mh_4}
  S(h_4,\vec z, \vec w) dh_4,
  \ee
  one obtains from (A~5.11) an equation  for $\bar S$,
  $$
  (-i\vec \gamma\vec \partial-im(1-\gamma_4)) \bar S(\vec z,
  \vec w) -i\sigma |\vec z| \{[1-\frac{1}{6}(1-\gamma_4)]+
  $$
  \be
  +[1-\frac{1}{6}(1+\gamma_4)]J(2mT_g)\}\bar S(\vec z, \vec
  w)= \delta^{(3)}(\vec z-\vec w)
  \ee
  where we have defined
  \be
  J(x)=exp(x^2)[1-\Phi(x)]=
  \left \{
  \begin{array}{ll}
  \frac{1}{\sqrt{\pi}x},&x\gg 1\\
  1-\frac{2x}{\sqrt{\pi}}+0(x^2),& x\ll 1
  \end{array}
  \right.
  \ee
  where $\Phi
  (x)$ is the error function.

  In arriving to (A~5.13) the change of variables
  $h_4-h'_4=u$, and integration over $dh_4 du$ helps to
  factorize out  $\bar S$ under the  integral.

  Till now we only have made an expansion in $1/m$ keeping the
  leading $0(m)$ and next-to-leading $0(m^0)$ terms, in particular
  the kernel $M\sim \sigma |\vec z|$ is $0(m^0)$ when one inserts in
  (A~5.7) $S_0$ instead of $S$, implying that $\sigma|\vec z|\ll m$.

  Therefore the product $mT_g$ is arbitrary and one can consider two
  limiting cases as in (A~5.14). For $mT_g\gg 1$ one can neglect the
  term with $J$ in (A~5.13) and obtains an equation for $\bar S=\left
  (
  \begin{array}{ll}
  S_{++}&S_{+-}\\
  S_{-+}&S_{--}\end{array}
  \right )
  $
  $$
  -\vec \sigma\vec\partial\bar S_{-+}-i\sigma|\vec z|\bar
  S_{++}=\delta^{(3)}(\vec z-\vec w)
  $$
  $$
  -i\sigma|\vec z|S_{+-}-\vec \sigma\vec\partial\bar S_{--}=0
  $$
  $$
  \vec \sigma\vec\partial\bar S_{++}- 2im
  \bar
  S_{-+}-\frac{2i}{3}\sigma
  |\vec z| \bar S_{-+}=0
  $$
  \be
  \vec \sigma\vec\partial\bar S_{+-}- 2im
  \bar
  S_{--}-i\frac{2}{3}\sigma
  |\vec z| \bar S_{--}=
  \delta^{(3)}(\vec z-\vec w)
  \ee

  To the leading order one obtains as expected
  \be
  (\frac{\vec p^2}{2m}+\sigma|\vec z|)\bar S_{++}=i\delta^{(3)}(\vec
  z-\vec w)
  \ee
  In the opposite limit, when $mT_g\ll 1$, one has from (A~5.13)
  \be
  [-i\vec \gamma\vec \partial-im (1-\gamma_4)]\bar
  S-i\frac{5}{3}\sigma|\bar z|\bar S=\delta^{(3)}(\vec z-\vec w)
  \ee
  and in leading order of $1/m$ one obtains instead of (A~5.16)
  \be
  (\frac{\vec p^2}{2m}+\frac{5}{3}\sigma|\vec z|)\bar
  S_{++}=i\delta^{(3)}(\vec z-\vec w)
  \ee

    We conclude this appendix by discussion of the role  of magnetic
  and electric correlators $D^B,D^E$ on the properties of confinement
  of the light quark.

  Having in mind that in (A~5.7) the term $\gamma_4S\gamma_4$
  enters with $J^E\sim D^E$, while $\gamma_i S\gamma_i$ is
  multiplied with $J^B\sim D^B$, one can see that for $D^E\equiv 0,
  D^B\neq 0$ one would have in (A~5.8)
  \be
  iM^B=\frac{i}{2}\tilde J^B(\vec z, \vec z) \delta^{(3)} (\vec
  z-\vec w)\{ \theta(h_4)
  e^{-mh_4}\frac{2}{3}(1-\gamma_4)+
   \theta(-h_4)
  e^{mh_4}\frac{2}{3}(1+\gamma_4)\}
  \ee
  To the leading order in $1/m$ one obtains the free equation instead
  of (A~5.16) implying that magnetic correlator $D^B$
  does not lead to confinement.

 \end{document}